\documentclass{IEEEtaes}
\usepackage{comment}
\usepackage{color,array,amsthm}
\usepackage{graphicx}
\usepackage{amsmath}
\usepackage{graphicx}
\usepackage{subcaption}
\usepackage{tikz}
\usetikzlibrary{shapes.geometric, arrows.meta, positioning, calc}

\begin{document}

\title{Physics Driven Digital Twin Model for Evaluation of GNSS User Receiver Equipment}

\author{Jitu Sanwale}
\affil{Indian Institute of Technology Kanpur, Kanpur–208 016, Uttar Pradesh, India}

\author{Mangal Kothari\footnote{Mangal Kothari contributed to this work while he was with the Department of Aerospace Engineering, Indian Institute of Technology Kanpur.}}
\member{Member, IEEE}
\affil{ADASI, EDGE Group, Abu Dhabi Emirate, United Arab Emirates}

\author{Hari B. Hablani}
\member{Senior Member, IEEE}
\affil{Indian Institute of Technology Indore, Indore–453 552, Madhya Pradesh, India}

\author{Suresh Dahiya}
\member{Member, IEEE}
\affil{Sardar Vallabhbhai National Institute of Technology Surat, Surat–395 007, Gujarat, India}

\corresp{Corresponding author: Jitu Sanwale, e-mail: sanwalejitu@gmail.com}

\maketitle

\begin{abstract}
This paper presents a physics-consistent digital twin framework for end-to-end modeling and evaluation of Global Navigation Satellite System (GNSS) user receiver equipment. In contrast to conventional GNSS simulators that rely on predefined signal models, the proposed framework enforces dynamic consistency between satellite ephemerides, user motion, and received signal observables through trajectory-driven injection of code-phase and Doppler dynamics. The GPS L1 C/A signal is synthesized in accordance with the IS-GPS-200 Rev.~N specification, with motion-induced effects derived directly from orbital and user kinematics, and augmented by ionospheric and tropospheric delay models. The resulting complex baseband signal is upconverted to radio frequency using a software-defined radio (SDR) platform disciplined by an external reference clock, enabling seamless hardware-in-the-loop (HIL) integration with commercial and software receivers. Validation across static, moderate-motion, and high-dynamics scenarios—including projectile-like trajectories—demonstrates close agreement between truth-model and receiver-estimated code phase, Doppler, and position, as well as strong correspondence between simulated and measured intermediate-frequency (IF) spectra. The results establish the proposed digital twin as a high-fidelity, repeatable, and physically consistent platform for GNSS receiver evaluation, tracking-loop stress testing, and development of robust navigation algorithms.
\end{abstract}


\section{INTRODUCTION}

Global Navigation Satellite Systems (GNSS) underpin a broad spectrum of aerospace, transportation, and timing applications in which reliable positioning, navigation, and timing (PNT) information is essential~\cite{morton2021position}. Verification and validation (V\&V) of GNSS receiver systems for mission-critical operations must satisfy stringent safety-of-life requirements on accuracy, availability, continuity, and integrity, as specified in standards such as RTCA DO-229, DO-253, and EUROCAE ED-95~\cite{unoosaIndex}. As modern receivers increasingly operate in dynamic, interference-prone, and multipath-rich environments, the need for high-fidelity and repeatable test methodologies has become more pronounced.

Traditional GNSS receiver testing relies on field trials, record-and-replay systems, and laboratory simulators~\cite{teunissen2017springer}. Field testing provides realistic conditions but suffers from poor repeatability and limited controllability. Record-and-replay systems enable controlled playback of real RF signals but offer limited flexibility in scenario design. Commercial GNSS simulators provide repeatability and configurability, yet they typically depend on predefined signal models that may not enforce physical consistency between satellite motion, user trajectory, atmospheric effects, and the resulting code-phase and Doppler observables. Consequently, they may not accurately reproduce the tightly coupled dynamics encountered in high-dynamics or contested environments.

Digital-twin technology has recently emerged as a powerful paradigm for modeling cyber-physical systems through physics-consistent virtual representations. In the GNSS domain, digital twins have been explored for applications such as multipath characterization~\cite{addo2024digital}, urban-positioning enhancement~\cite{lian2024improving}, and spoofing-robustness evaluation using twin-based receiver replicas~\cite{danelli2026gnss}. These studies demonstrate the growing relevance of digital-twin methodologies; however, existing efforts are typically application-specific and do not provide an end-to-end framework that simultaneously models satellite dynamics, propagation effects, and receiver processing in a physically consistent manner. 

Recent publications highlight receiver algorithms and designs that address spoofing, complex multipath environments, and high-dynamics scenarios \cite{venturino2025adaptive, xu2023deeply, won2012iterative}. They also emphasize the importance of timing, integrity, and system-level GNSS receiver behavior~\cite{anderson2025tesla,blanch2025gaussian}. Together, these works underscore the necessity of test frameworks that preserve physical coherence throughout the entire signal chain.

Accurate reproduction of the GNSS signal chain—from satellite transmission and propagation through the ionosphere and troposphere to receiver acquisition and tracking—is essential for rigorous receiver evaluation. This requires consistent modeling of code-phase evolution, Doppler shifts, and propagation delays. The challenge becomes particularly acute for high-dynamics platforms such as unmanned aerial vehicles, guided munitions, and projectiles, where rapid variations in Doppler and Doppler rate can severely stress receiver tracking loops~\cite{ramteke2025dualspin, dahiya2022spinning}. These scenarios demand simulation frameworks that preserve the physical coupling between orbital dynamics, user motion, and received signal characteristics.

Motivated by these requirements, this paper presents a physics-driven digital-twin framework for GNSS signal generation and receiver evaluation. The proposed approach synthesizes complex baseband signals by injecting trajectory-consistent code-phase delays and Doppler dynamics derived directly from satellite ephemerides and user kinematics. This ensures that the generated signal remains physically consistent with the expected receiver observables, enabling rigorous assessment of acquisition, tracking, and navigation performance.

The synthesized signal is upconverted to radio frequency (RF) using a software-defined radio (SDR) platform, enabling hardware-in-the-loop (HIL) testing with both commercial and software GNSS receivers. The framework incorporates ionospheric and tropospheric delay models to emulate realistic propagation conditions. Validation is performed across static, moderate-motion, and high-dynamics scenarios, demonstrating strong agreement between truth-model observables and receiver-estimated code phase, Doppler, and position, as well as close correspondence between simulated and measured intermediate-frequency (IF) spectra.

The main contributions of this work are:
\begin{itemize}
    \item A physics-consistent end-to-end GNSS digital twin integrating satellite dynamics, propagation effects, and receiver processing.
    \item A trajectory-driven signal synthesis method that injects code-phase and Doppler dynamics consistent with satellite ephemerides and user motion.
    \item A hardware-in-the-loop validation architecture using SDR platforms, enabling seamless transition between simulated IF and radiated RF signals.
    \item A reproducible stress-testing framework for GNSS tracking loops under static, moderate-motion, and high-dynamics conditions.
    \item Quantitative validation of receiver performance through comparison of truth-model and receiver-estimated observables, including code phase, Doppler, pseudorange, and position.
\end{itemize}

The remainder of this paper details the signal generation model, propagation channel representation, user motion dynamics, and validation experiments demonstrating the fidelity and robustness of the proposed digital twin.

\section{DIGITAL TWIN FRAMEWORK}

\begin{figure}
    \centering
    \includegraphics[width=1\linewidth]{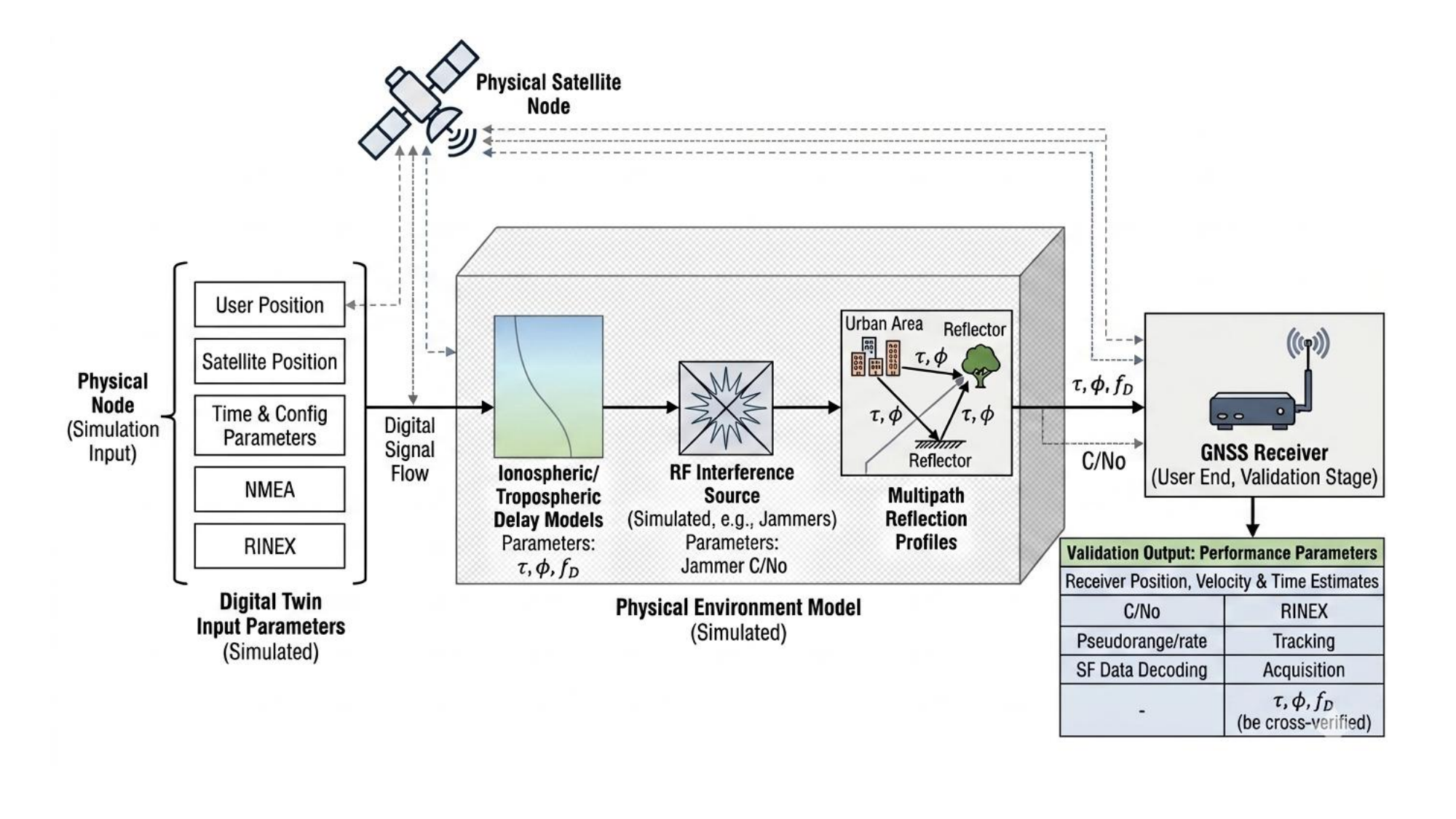}
    \caption{ Physics-driven digital twin framework for GNSS signal propagation and receiver validation}
    \label{fig:digitalTwin-sg}
\end{figure}

The overall architecture of the proposed digital twin framework is shown in Fig.~\ref{fig:digitalTwin-sg}. The digital twin models the end-to-end satellite–receiver signal chain through three tightly coupled components: (i) a physics-based GNSS signal simulation model, (ii) a relative motion model that governs code-phase and Doppler dynamics, and (iii) a received signal power model that determines the carrier-to-noise density ratio ($C/N_0$). These components explicitly incorporate satellite and receiver antenna patterns, free-space path loss, and propagation effects. External impairments—including interference and multipath—are also modeled to emulate realistic operating conditions. The measurement models for ionospheric and tropospheric delays, described in Sec.~\ref{PropModel}, provide physically consistent propagation delays that integrate with the simulation model.

\subsection{Simulation Model}

The simulation model digitally synthesizes the navigation signal transmitted by a GNSS satellite, preserving its spectral, modulation, and timing characteristics.

\subsubsection{GNSS Signal Model}
\label{SigModel}

The transmitted signal is modeled following \cite{morton2021position} as
\begin{equation} \label{TxSign}
\begin{split}
s_{TX}(t) &= \sqrt{2P_{TX,I}} 
\cos\!\left[ 2\pi\!\left(f_{TX}+\delta f_{SV}(t)\right)t + \phi_I(t) \right] \\
&\quad + \sqrt{2P_{TX,Q}}
\sin\!\left[ 2\pi\!\left(f_{TX}+\delta f_{SV}(t)\right)t + \phi_Q(t) \right],
\end{split}
\end{equation}
where $P_{TX,I}$ and $P_{TX,Q}$ denote the in-phase and quadrature transmit powers, $f_{TX}$ is the nominal carrier frequency, $\delta f_{SV}(t)$ is the satellite oscillator drift, and $\phi_I(t)$ and $\phi_Q(t)$ are the instantaneous phases.

For binary phase-quadrature modulation, the phase terms are
\begin{equation} \label{phi-I}
\phi_I(t) = \pi\!\left\{ D_I(F_D,t) \bigoplus C_I(\bar{f}_C,t) \right\},
\end{equation}
\begin{equation} \label{phi_Q}
\phi_Q(t) = \pi\!\left\{ D_Q(F_D,t) \bigoplus C_Q(f_C,t) \right\},
\end{equation}
where $D$ is the navigation data bit, $C_I$ and $C_Q$ are the PRN codes, and $\bigoplus$ denotes the XOR operation.

Modern signals employ secondary codes, yielding
\begin{equation} \label{phi-I1}
\phi_I(t) = \pi\!\left\{ D_I \bigoplus C_I \bigoplus C_{I,SC} \right\},
\end{equation}
\begin{equation} \label{phi_Q1}
\phi_Q(t) = \pi\!\left\{ D_Q \bigoplus C_Q \bigoplus C_{Q,SC} \right\},
\end{equation}
where $C_{I,SC}$ and $C_{Q,SC}$ are secondary codes generated at rate $f_{SC}$.

After propagation through the ionosphere and troposphere, the received signal becomes
\begin{equation} \label{Rx}
\begin{split}
s_{RX}(t) &= \sqrt{P_{RX,I}} 
\cos\!\left[ \omega_{RX} t + \phi_I(t-\tau(t)) + \phi_{RX} \right] \\
&\quad + \sqrt{P_{RX,Q}}
\sin\!\left[ \omega_{RX} t + \phi_Q(t-\tau(t)) + \phi_{RX} \right],
\end{split}
\end{equation}
where $\omega_{RX}=2\pi(f_{TX}+\delta f_{SV}(t)+f_D(t))$ and $\tau(t)$ is the total propagation delay.

The pseudorange is modeled as \cite{enge2012global}
\begin{equation}
\tau = \frac{\|\mathbf{r}_s - \mathbf{r}_u\|}{c}
+ \frac{I_\rho(t)}{c}
+ \frac{T_\rho(t)}{c},
\end{equation}
where $\mathbf{r}_s$ and $\mathbf{r}_u$ are satellite and user positions, and $I_\rho$ and $T_\rho$ are ionospheric and tropospheric delays.

\subsubsection{User Receiver Motion Model}
\label{userMotion}

\begin{figure}[t]
\centering
\includegraphics[width=8.5cm]{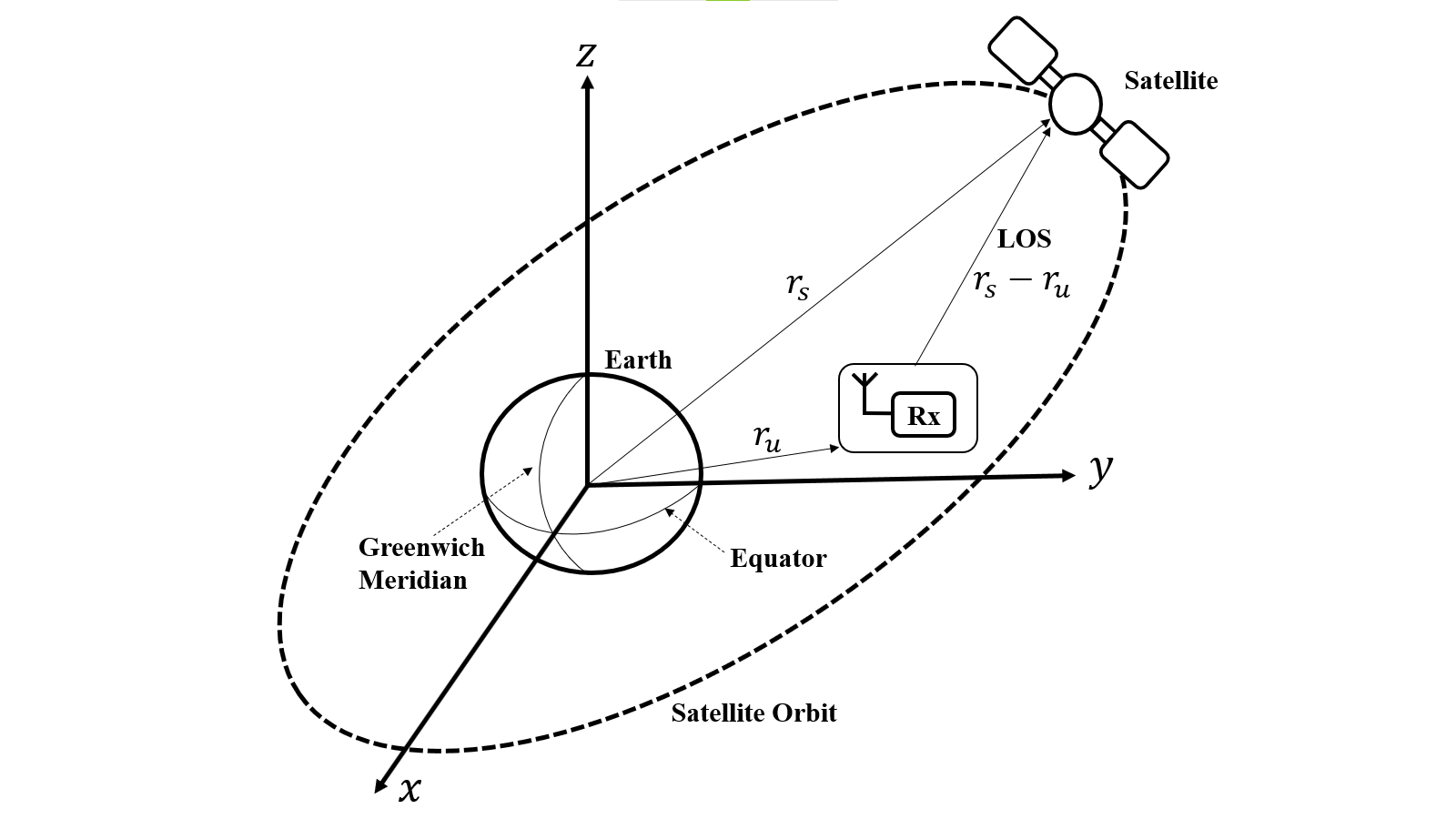}
\caption{Satellite–receiver geometry in the ECEF frame.}
\label{LOS}
\end{figure}

Figure~\ref{LOS} shows the relative geometry between the satellite and the user receiver in the Earth-Centered, Earth-Fixed (ECEF) frame. This geometric relationship forms the basis for modeling how the received signal evolves as either platform moves.

Because the satellite and user are in continuous relative motion, the received carrier experiences a Doppler shift. This shift is determined by the projection of the relative velocity vector onto the line-of-sight (LOS) direction and is expressed as
\begin{equation} \label{Doppler}
f_D = -\frac{f_{TX}}{c}
\frac{(\mathbf{v}_s - \mathbf{v}_u)\cdot(\mathbf{r}_s - \mathbf{r}_u)}
{\|\mathbf{r}_s - \mathbf{r}_u\|},
\end{equation}
where $\mathbf{v}_s$ and $\mathbf{v}_u$ denote the satellite and user velocities, and $\mathbf{r}_s$ and $\mathbf{r}_u$ are their respective position vectors. This formulation ensures that the Doppler dynamics injected into the synthesized signal remain physically consistent with the underlying orbital and user motion.

\subsubsection{Received Signal Power and $C/N_0$ Modeling}
\label{CNR}

The strength of the signal received from satellite $k$ depends on the satellite–receiver geometry, antenna characteristics, and free-space attenuation. A convenient way to capture these effects is through the Friis transmission relationship,
\begin{equation}
C_k = P_{TX,k} \, G_{TX,k}(\theta_k) \, G_{RX}(\theta_k)
\left( \frac{\lambda}{4\pi R_k} \right)^2
L_{\mathrm{atm},k},
\label{eq:friis}
\end{equation}
where $R_k$ is the satellite–receiver range and $\theta_k$ is the elevation angle. The term $L_{\mathrm{atm},k}$ accounts for atmospheric and implementation losses, while the antenna gains reflect the directional response of both the satellite and receiver antennas. For simulation purposes, the constant factors in \eqref{eq:friis} can be grouped, leading to the simplified dependence
\begin{equation}
C_k \propto \frac{G(\theta_k)}{R_k^2},
\label{eq:power_simplified}
\end{equation}
where $G(\theta_k)$ represents the combined elevation-dependent gain. The corresponding carrier-to-noise density ratio is then
\begin{equation}
\left( \frac{C}{N_0} \right)_k = \frac{C_k}{N_0},
\label{eq:cn0_def}
\end{equation}
and the baseband signal for satellite $k$ is scaled by
\begin{equation}
A_k = \sqrt{C_k}.
\label{eq:amplitude}
\end{equation}

The composite received signal is formed by summing the contributions from all visible satellites and adding complex additive white Gaussian noise (AWGN) with variance $\sigma^2 = N_0 f_s$. This ensures that variations in $C/N_0$ arise naturally from geometry and antenna effects, while thermal noise remains consistent across channels.

\subsubsection{Interference Modeling}

GNSS receivers routinely encounter a variety of radio-frequency interference sources, and a realistic digital twin must be able to reproduce these effects. The framework therefore includes several commonly observed interference types, each modeled in a form suitable for signal-level simulation.

\begin{itemize}
\item \textbf{Chirp interference:}  
A chirp signal sweeps across a range of frequencies and can overlap a significant portion of the GNSS band, making it particularly disruptive to acquisition and tracking. It is modeled as  
\(
i(t)=A e^{j(2\pi\int_0^t f(\tau)d\tau + \phi)}, \quad f(t)=f_0+kt,
\)  
where the instantaneous frequency increases linearly with rate $k$.

\item \textbf{Continuous-wave interference (CWI):}  
CWI represents a narrowband tone at a fixed frequency. Even though it occupies little bandwidth, it can desensitize the receiver front end or bias carrier tracking loops. It is expressed as  
\(
i(t)=A e^{j(2\pi f_i t + \phi_i)}.
\)

\item \textbf{FMCW interference:}  
Frequency-modulated continuous-wave signals introduce a rapidly varying instantaneous frequency, often producing periodic distortions in the correlator outputs. The model used is  
\(
i(t)=A e^{j(2\pi f_i t + \beta\sin(2\pi f_m t))},
\)  
where $\beta$ and $f_m$ control the modulation depth and rate.

\item \textbf{Pulse interference:}  
Short-duration, high-power RF pulses can momentarily overwhelm the receiver front end. Their impact depends on pulse width, repetition rate, and spectral content, and they are included to emulate environments with impulsive or burst-like interference.
\end{itemize}

By incorporating these interference types, the digital twin enables controlled evaluation of receiver robustness under a range of realistic RF conditions, from benign to highly stressed scenarios.

\subsubsection{Multipath Modeling}

Multipath arises when the GNSS signal reaches the receiver not only through the direct line-of-sight (LOS) path but also through reflections from surrounding surfaces such as buildings, terrain, or water. These reflected components introduce additional delays and phase distortions, and are a major source of ranging error, particularly in urban or cluttered environments. To capture these effects, the received signal is modeled following \cite{kaplanHegarty} as
\begin{equation}
\overline{s}_{RX}(t) = s_{RX}(t-\tau_0)
+ \sum_{i=1}^{M} \alpha_i
e^{j[\omega_{RX} t + \phi_I(t-\tau_i(t)) + \phi_{RX}]},
\end{equation}
where $\tau_0$ is the LOS delay, and each reflected path is characterized by an amplitude $\alpha_i$ and an excess delay $\tau_i(t)$. This formulation allows the digital twin to reproduce both mild and severe multipath conditions by adjusting the number, strength, and delay spread of the reflected components.

\subsection{Measurement Model}
\label{PropModel}

The measurement model provides physically consistent ionospheric and tropospheric delays using parameters derived from broadcast navigation data and near–real-time GNSS products \cite{RTS-JPL}. These delays are incorporated directly into the signal generation process, ensuring that the synthesized observables reflect realistic propagation conditions.

\subsubsection{Ionosphere}

The ionosphere introduces several frequency-dependent effects on GNSS signals, most notably group delay and Faraday rotation \cite{ITU-R}. For L-band signals, group delay is the dominant contributor to ranging error, while other effects are typically small enough to be neglected. A variety of ionospheric models exist—such as NeQuick, IRI, and grid-based TEC maps—but for real-time applications the broadcast Klobuchar model remains widely used due to its simplicity and low computational cost.

In this work, ionospheric delay is modeled using the Klobuchar formulation \cite{4104345}, with the coefficients
\(
\alpha_0,\ldots,\alpha_3,\;
\beta_0,\ldots,\beta_3
\)
taken directly from the navigation message. Although approximate, the model captures the first-order diurnal variation of the ionosphere and provides delay estimates consistent with operational GPS receivers.

\subsubsection{Troposphere}

Unlike the ionosphere, the troposphere introduces a nondispersive delay that affects all GNSS frequencies equally. The delay depends primarily on surface pressure, temperature, and humidity, and is commonly separated into hydrostatic and wet components. In this work, the zenith tropospheric delay is computed using the Saastamoinen model \cite{saastamoinen1973contributions},
\begin{equation} \label{Saastamoinen}
ZTD = 0.002277
\frac{P_0 + \left(0.05 + \frac{1255}{T_0+273.15}\right)\chi}
{1 - 0.00266\cos(2\varphi) - 0.00028h},
\end{equation}
where $P_0$, $T_0$, $RH$, $\chi(= RH \times 6.11 \times 10^{\frac{7.5 T_0}{T_0 + 273.3}})$, $\varphi$, and $h$ denote the surface pressure, temperature, relative humidity, water vapor pressure indicator, latitude, and height, respectively.

To map the zenith delay to the actual line-of-sight direction, standard elevation-dependent mapping functions—such as Niell, VMF3, or GPT3—are applied. These functions allow the digital twin to reproduce realistic slant delays for satellites at different elevation angles, ensuring consistency with real-world GNSS observations.

\section{EXPERIMENTAL RESULTS AND DISCUSSION}

This section validates the proposed physics-driven digital twin using both software-based simulations and hardware-in-the-loop (HIL) experiments. All experiments are performed for the GPS L1 C/A signal. The objective is to demonstrate that the synthesized signal reproduces physically consistent code-phase, Doppler, and position observables across a range of user motion profiles. Validation is carried out through: (i) analysis of the power spectral density (PSD) of intermediate-frequency (IF) samples, (ii) comparison of truth-model observables with those estimated by a GNSS software receiver, and (iii) comparison of position solutions obtained from a software receiver and a commercial GNSS module in HIL mode.
\usetikzlibrary{shapes.geometric, arrows.meta, positioning, decorations.pathreplacing}
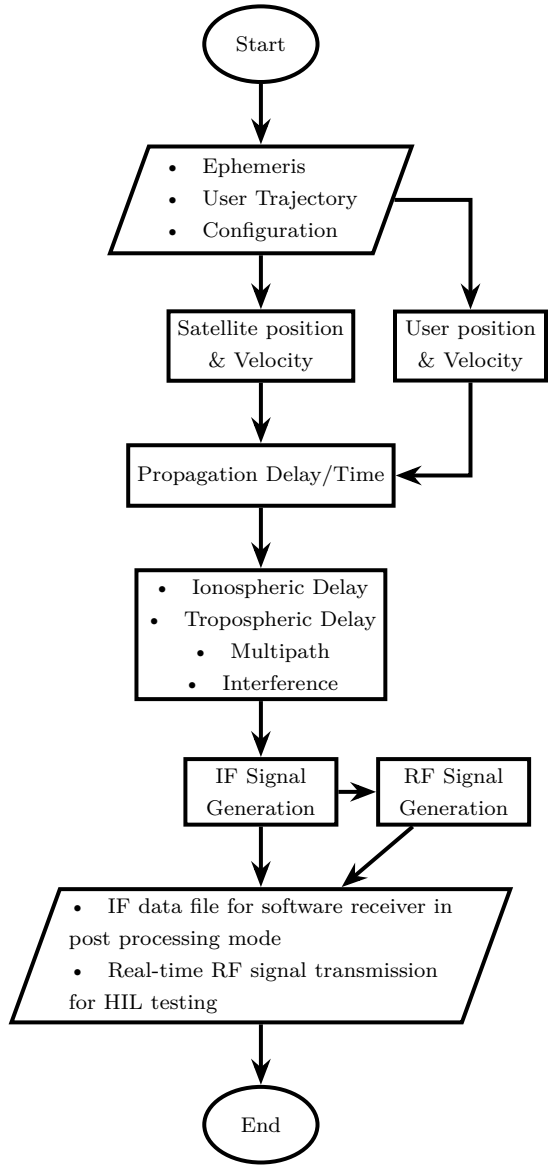
\begin{figure}
 \centering
\begin{tikzpicture}[
    font=\sffamily\small\bfseries,
    node distance=0.8cm and 1cm,
    terminal/.style={ellipse, draw, line width=1.5pt, minimum width=1.5cm, minimum height=1cm},
    process/.style={rectangle, draw, line width=1.5pt, minimum width=2cm, minimum height=0.8cm, align=center},
    io/.style={trapezium, trapezium left angle=70, trapezium right angle=110, draw, line width=1.5pt, minimum width=4cm, minimum height=1cm, align=left},
    arrow/.style={-Stealth, line width=1.5pt}]

    \node [terminal] (start) {Start};

    \node [io, below=of start] (input1) 
    {
        \textbullet\quad Ephemeris\\
        \textbullet\quad User Trajectory\\
        \textbullet\quad Configuration
    };

    \path (input1.south) ++(0,-1.2cm) node[process] (sat_pos)    
    {Satellite position\\\& Velocity};

    \node [process, right=0.5cm of sat_pos] (user_pos) 
    {User position\\\& Velocity};

    \node [process, below=0.8cm of sat_pos] (prop_delay) 
    {Propagation Delay/Time};

    \node [process, below=of prop_delay, minimum height=1.5cm] (delays) 
    {
        \textbullet\quad Ionospheric Delay\\
        \textbullet\quad Tropospheric Delay\\
        \textbullet\quad Multipath \\
        \textbullet\quad Interference
    };

    \path (delays.south) ++(0,-1.2cm) node[process] (if_gen)    
     {IF Signal \\ Generation};

    \node [process, right=0.5cm of if_gen] (rf_gen) 
    {RF Signal \\ Generation};

     \node [io, below=0.8cm of if_gen] (output) 
    {
        \textbullet\quad IF data file for software receiver in \\post processing mode\\
        \textbullet\quad Real-time RF signal transmission \\for HIL testing
        
    };

    \node [terminal, below=of output] (end) {End};

    \draw [arrow] (start) -- (input1);
    \draw [arrow] (input1.south) -- (sat_pos.north);
    \draw [arrow] (input1.east) -| (user_pos.north);
    \draw [arrow] (sat_pos.south) -- (prop_delay.north);
    \draw [arrow] (user_pos.south) |- (prop_delay.east);
    \draw [arrow] (prop_delay) -- (delays);
    \draw [arrow] (delays) -- (if_gen);
    \draw [arrow] (if_gen) -- (rf_gen);
    \draw [arrow] (if_gen) -- (output.north);
    \draw [arrow] (rf_gen) -- (output);
    \draw [arrow] (output.south) -- (end);

    \end{tikzpicture}
        \caption{Workflow for virtual GNSS signal generation, including scenario definition, propagation delay estimation, and IF/RF signal synthesis.}
    \label{fig:digitalTwin}
\end{figure}

\subsection{Experimental Setup}

The digital twin generates GPS L1 C/A complex baseband samples at a configurable sampling rate and upconverts them to RF using a software-defined radio (SDR) front end. In this work, a HackRF One SDR is used due to its wide operating range (1~MHz–6~GHz), 8-bit quadrature sampling up to 20~Msps, and software-configurable gain and filtering. To ensure carrier-frequency stability, the SDR is disciplined by an external 10~MHz reference clock. The radiated or cabled RF signal is captured by either a commercial GNSS receiver or a custom software receiver.

The virtual signal generation pipeline (Fig.~\ref{fig:digitalTwin}) consists of three stages: (i) scenario generation using satellite ephemerides and user trajectory, (ii) estimation of ionospheric and tropospheric delays using near-real-time (NRT) measurements or RINEX-derived Klobuchar coefficients, and (iii) generation of IF samples incorporating motion-induced Doppler and code-phase dynamics. The resulting IF file is processed by a software receiver for acquisition, tracking, and navigation, and is also transmitted through the HackRF One for HIL testing with a SkyTraq PX1125S-01A receiver.

\begin{figure*}[t]
    \centering
    \includegraphics[width=\textwidth]{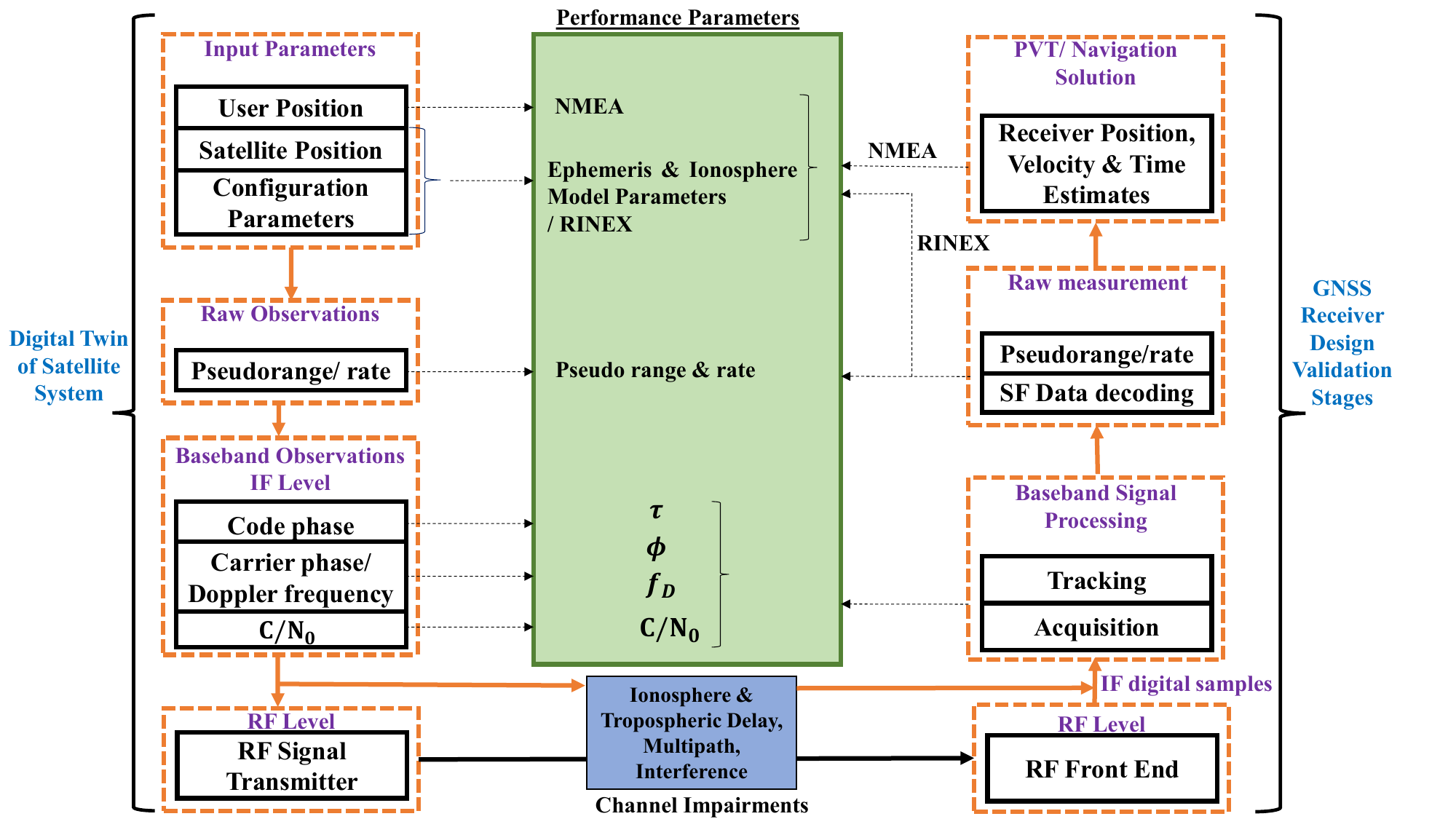}
    \caption{GNSS receiver verification and validation pipeline, including PSD estimation, acquisition, tracking, observable comparison, and PVT evaluation.}
    \label{fig:verification}
\end{figure*}

The validation pipeline shown in Fig.~\ref{fig:verification} includes:
\begin{enumerate}
    \item Estimation of PSD from recorded IF samples.
    \item Satellite acquisition and estimation of initial code and carrier parameters.
    \item Tracking by code and carrier loops and estimation of $C/N_0$.
    \item Generation of truth-model Doppler and code-phase based on ephemerides and user trajectory.
    \item Comparison of pseudorange, pseudorange rate, and carrier phase with truth-model observables.
    \item Navigation data decoding and verification against truth-model ephemerides.
    \item Position, velocity, and time (PVT) estimation and comparison with truth-model values.
\end{enumerate}

Three user-motion scenarios are considered:
\begin{itemize}
    \item \textbf{Static:} Receiver remains stationary to validate baseline Doppler stability, code-phase consistency, and navigation accuracy.
    \item \textbf{Moderate motion:} Representative of automotive, maritime, and UAV applications with velocities up to 20–30~m/s.
    \item \textbf{High dynamics:} Representative of rockets, missiles, and projectiles, with accelerations exceeding 20~g and Doppler rates above 50~Hz/s.
\end{itemize}

\begin{figure}[t]
    \centering
    \includegraphics[width=1\linewidth]{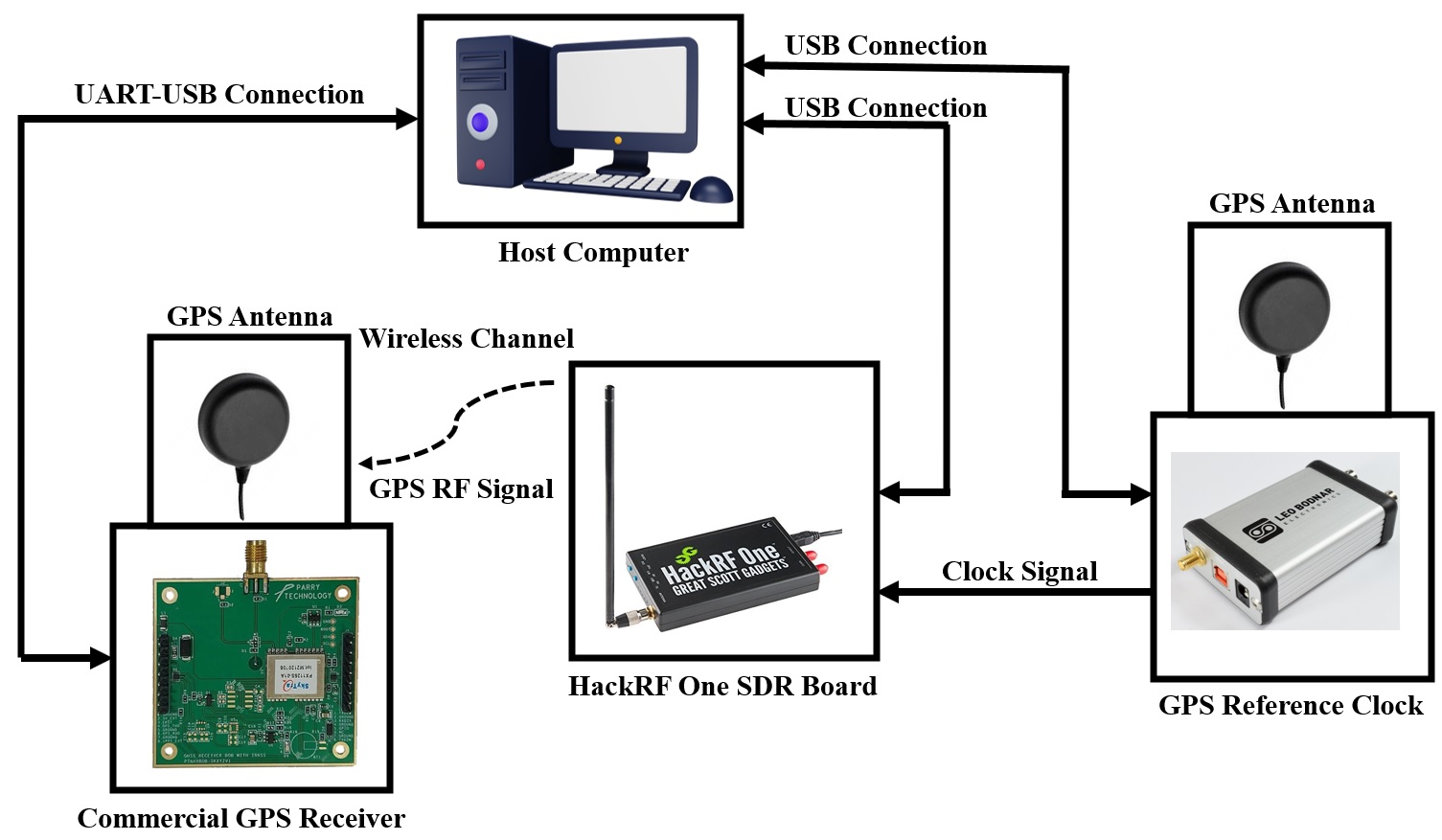}
    \caption{Hardware-in-the-loop (HIL) test configuration using HackRF One and SkyTraq PX1125S-01A receiver.}
    \label{fig:HIL}
\end{figure}

The HIL setup is shown in Fig.~\ref{fig:HIL}, where the host computer configures the reference clock, SDR, and GNSS evaluation board. A wireless link is used for RF transmission to the commercial receiver.

\subsection{Results and Discussion}

\begin{figure}[t]
    \centering
    \includegraphics[width=1\linewidth]{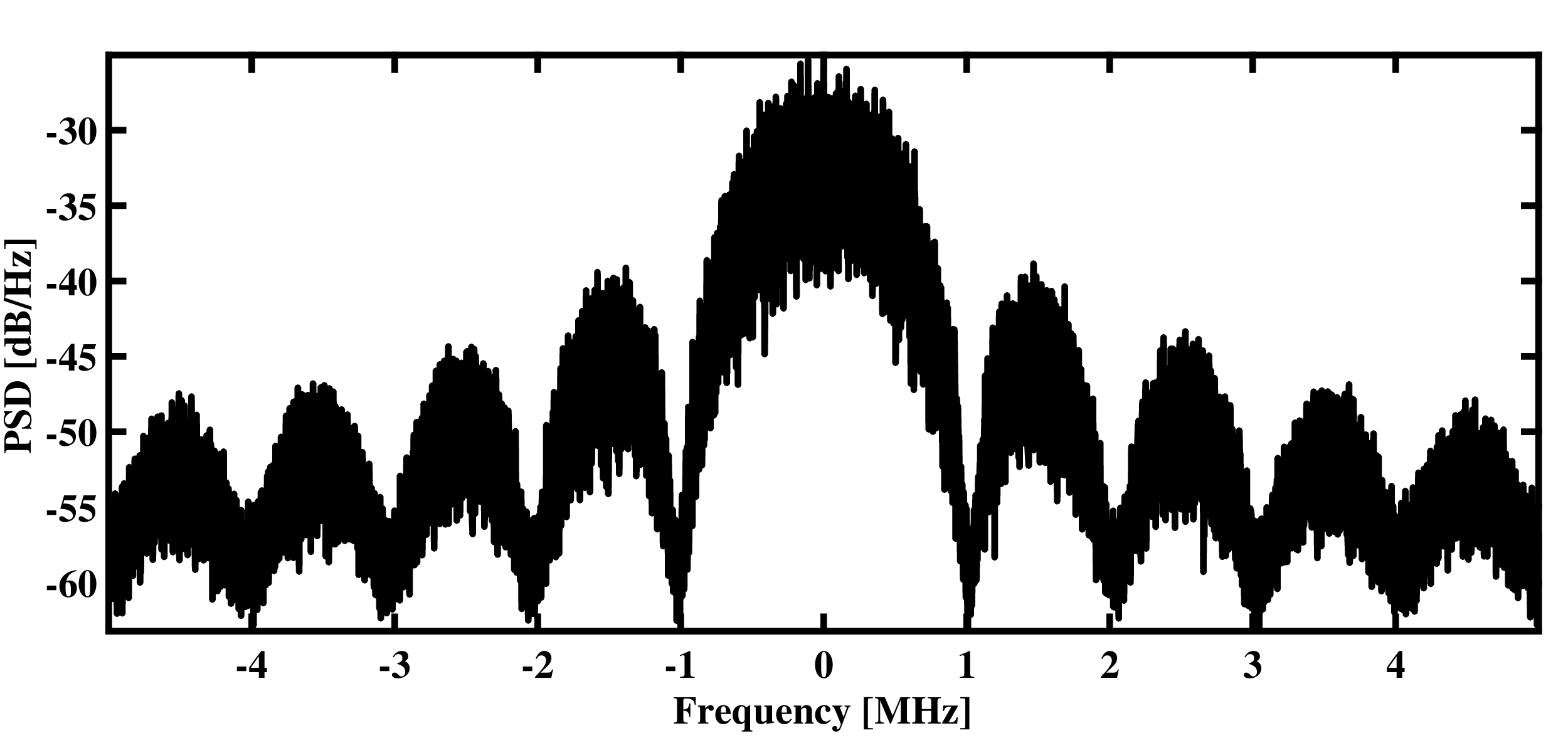}
    \caption{Power spectral density of the synthesized GPS L1 C/A signal, confirming correct BPSK(1) spectral characteristics.}
    \label{fig:psd}
\end{figure}

Experiments were carried out for both static and high‑dynamics user scenarios to evaluate the fidelity of the proposed digital twin. The first level of validation examines the signal domain. Figure~\ref{fig:psd} shows the PSD of the synthesized GPS L1 C/A signal. The expected sinc‑shaped spectrum of the BPSK(1) modulation, with nulls spaced at $\pm 1.023$~MHz around the IF frequency, a 0~MHz in our case, is clearly reproduced. This confirms that the digital twin preserves the correct chip rate, modulation characteristics, and spectral shape.

\begin{figure}[t]
    \centering
    \includegraphics[width=1\linewidth]{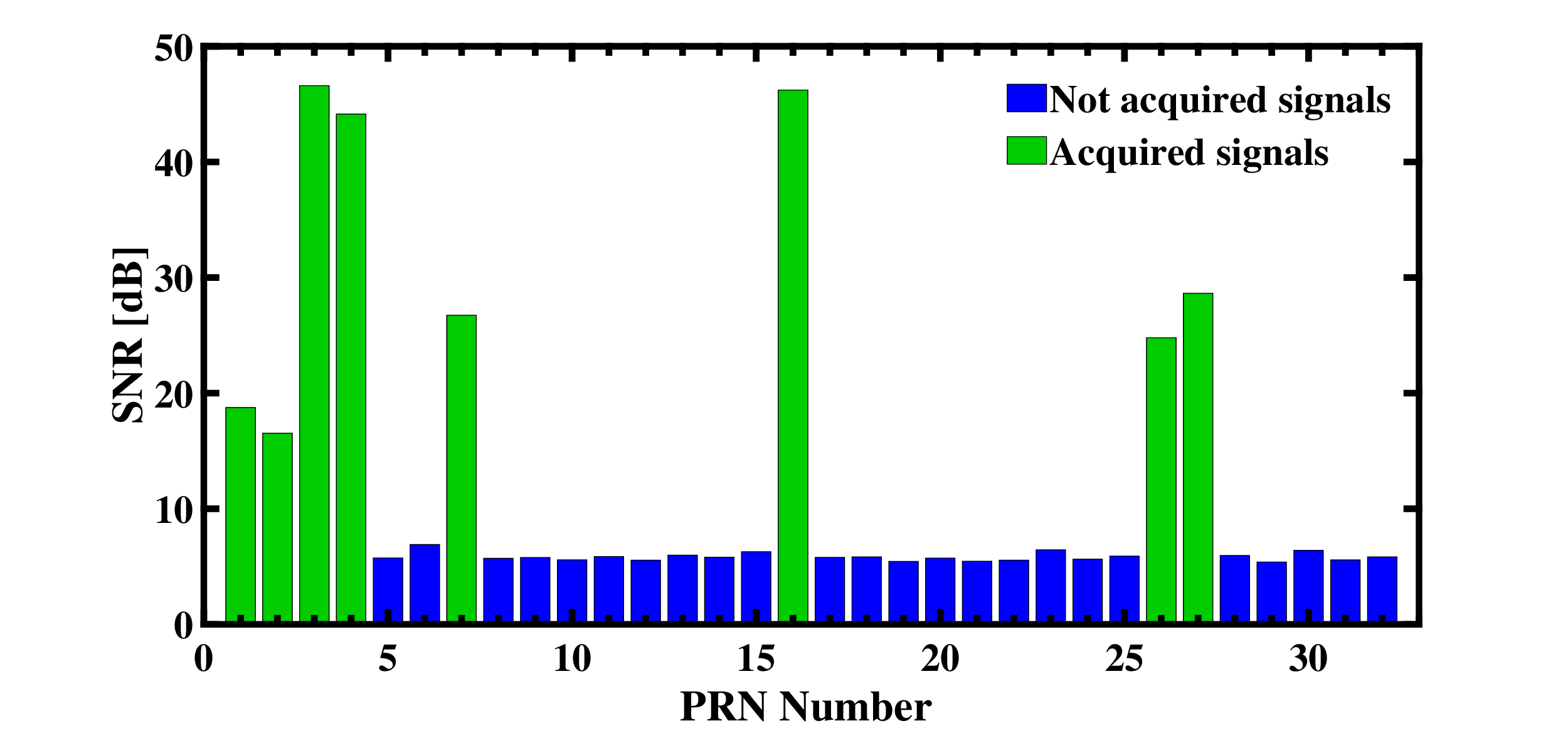}
    \caption{Acquisition results for the static scenario. Green bars indicate successfully acquired satellites.}
    \label{fig:acqResults}
\end{figure}

The acquisition results for the static scenario are shown in Fig.~\ref{fig:acqResults}. Satellites highlighted in green correspond to those present in the RINEX dataset used to generate the truth model. The match between the acquired PRNs and the truth data verifies correct PRN code generation and Doppler embedding in the synthesized signal.

\begin{figure*}[t]
    \centering
    \begin{subfigure}{0.3\linewidth}
        \includegraphics[width=\linewidth]{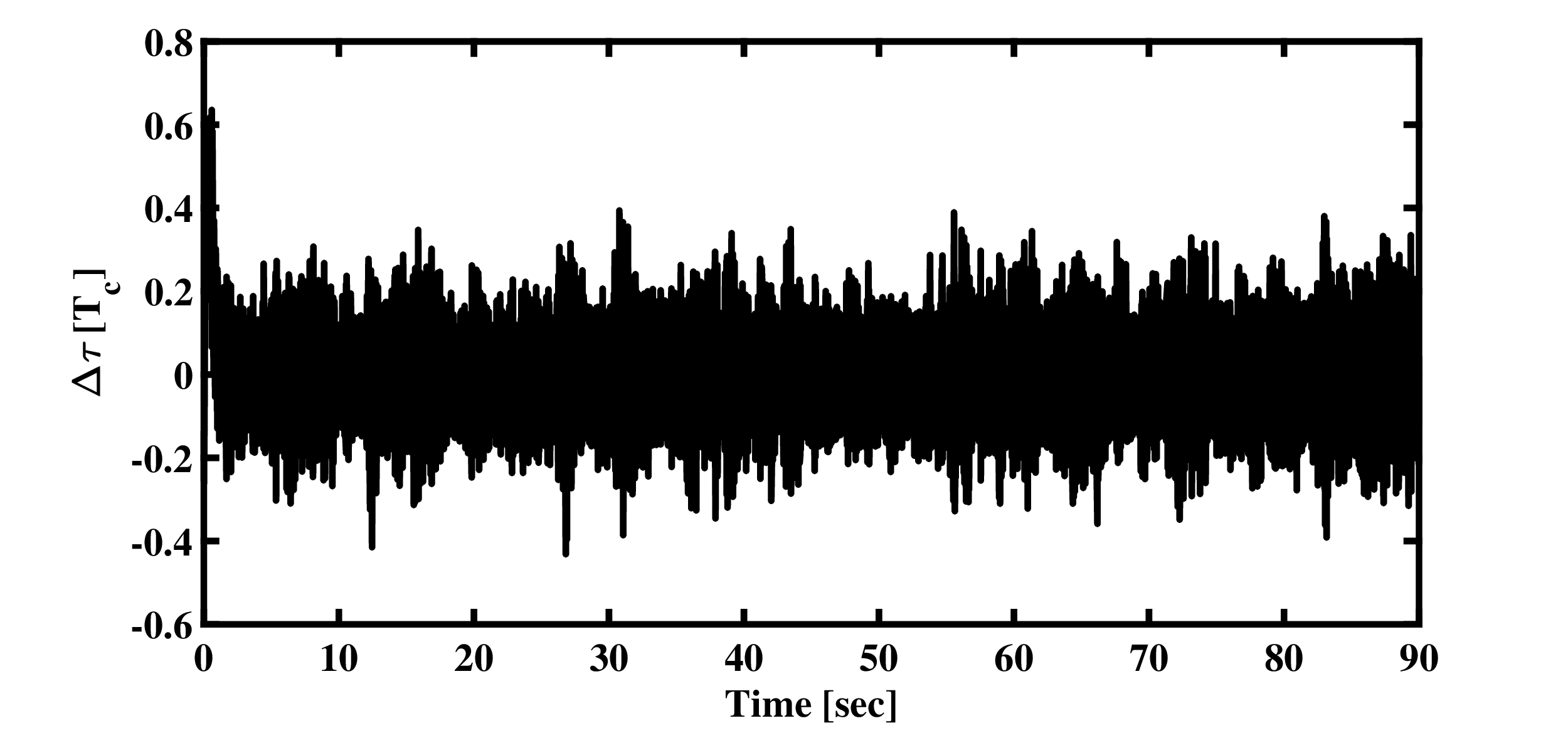}
        \caption{Code phase error (static)}
    \end{subfigure}
    \begin{subfigure}{0.3\linewidth}
        \includegraphics[width=\linewidth]{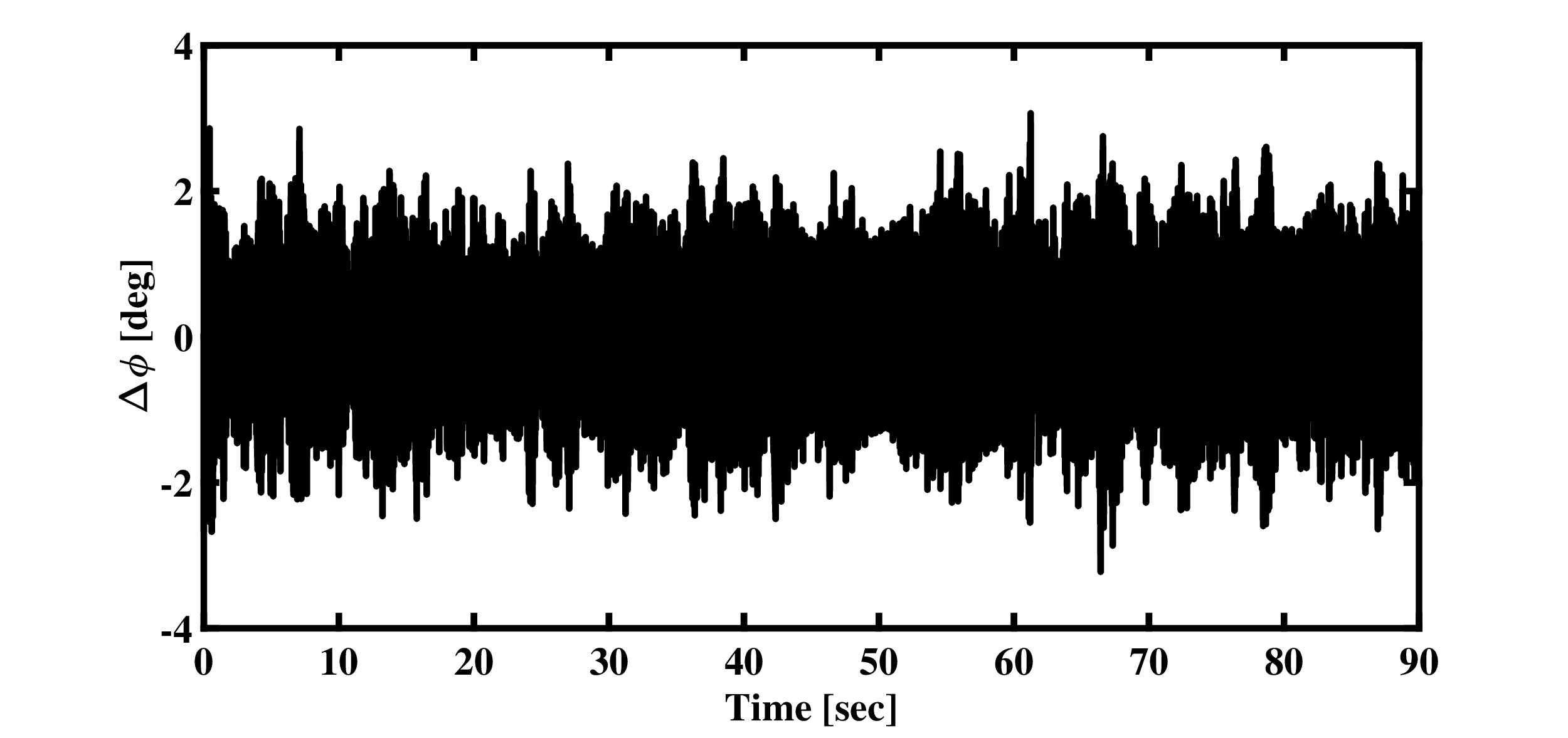}
        \caption{Carrier phase error (static)}
    \end{subfigure}
    \begin{subfigure}{0.3\linewidth}
        \includegraphics[width=\linewidth]{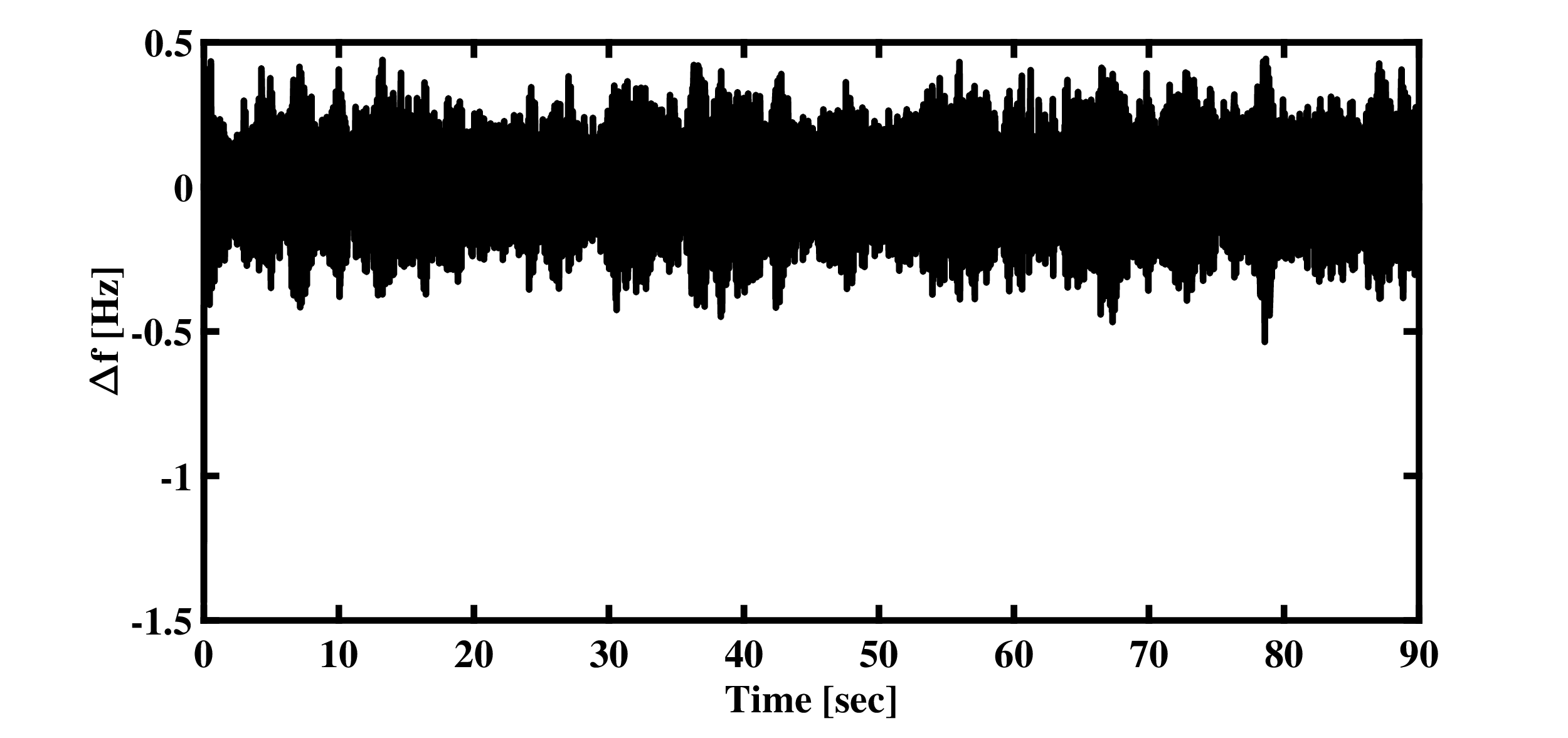}
        \caption{Carrier frequency error (static)}
    \end{subfigure}
    \vfill
    \begin{subfigure}{0.3\linewidth}
        \includegraphics[width=\linewidth]{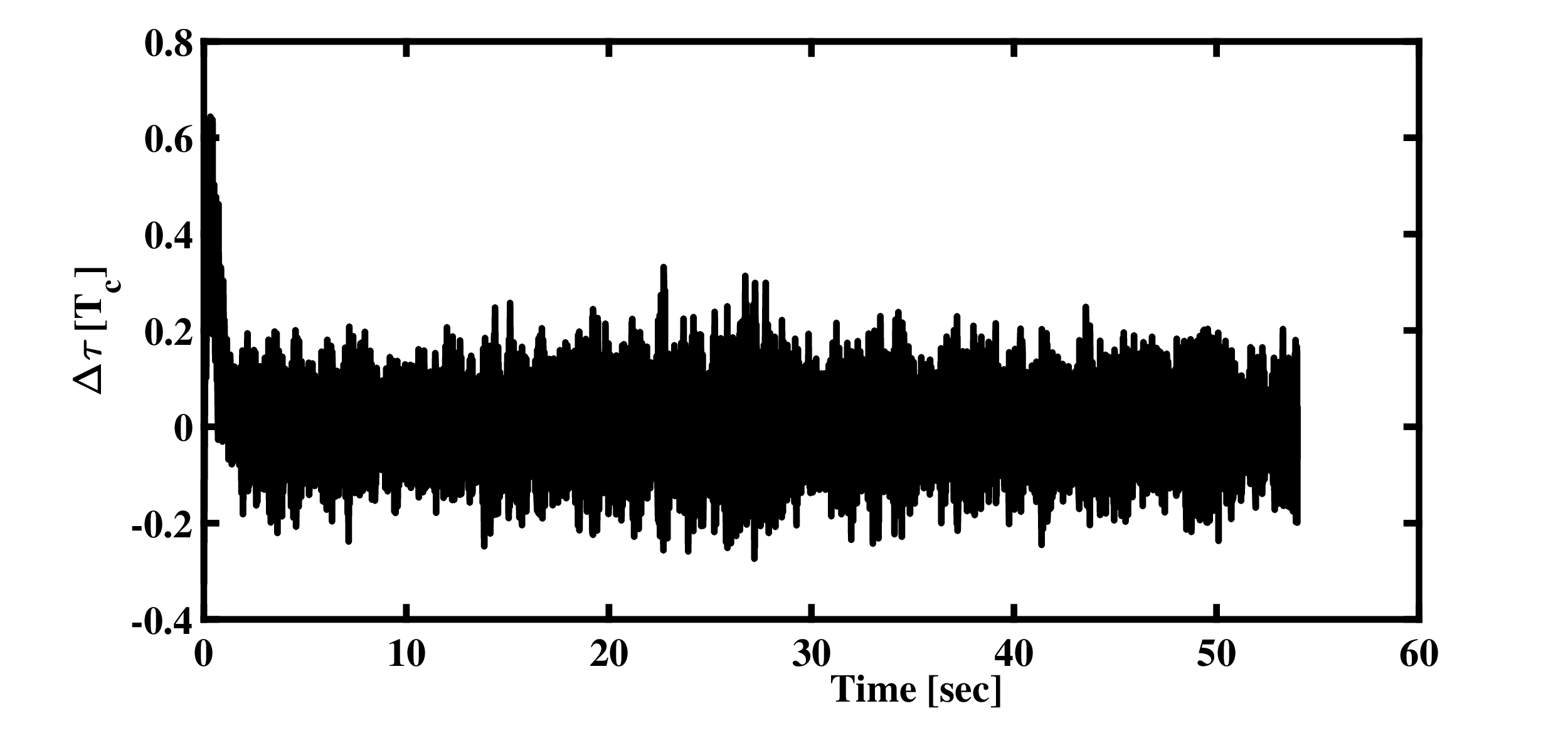}
        \caption{Code phase error (moderate)}
    \end{subfigure}
    \begin{subfigure}{0.3\linewidth}
        \includegraphics[width=\linewidth]{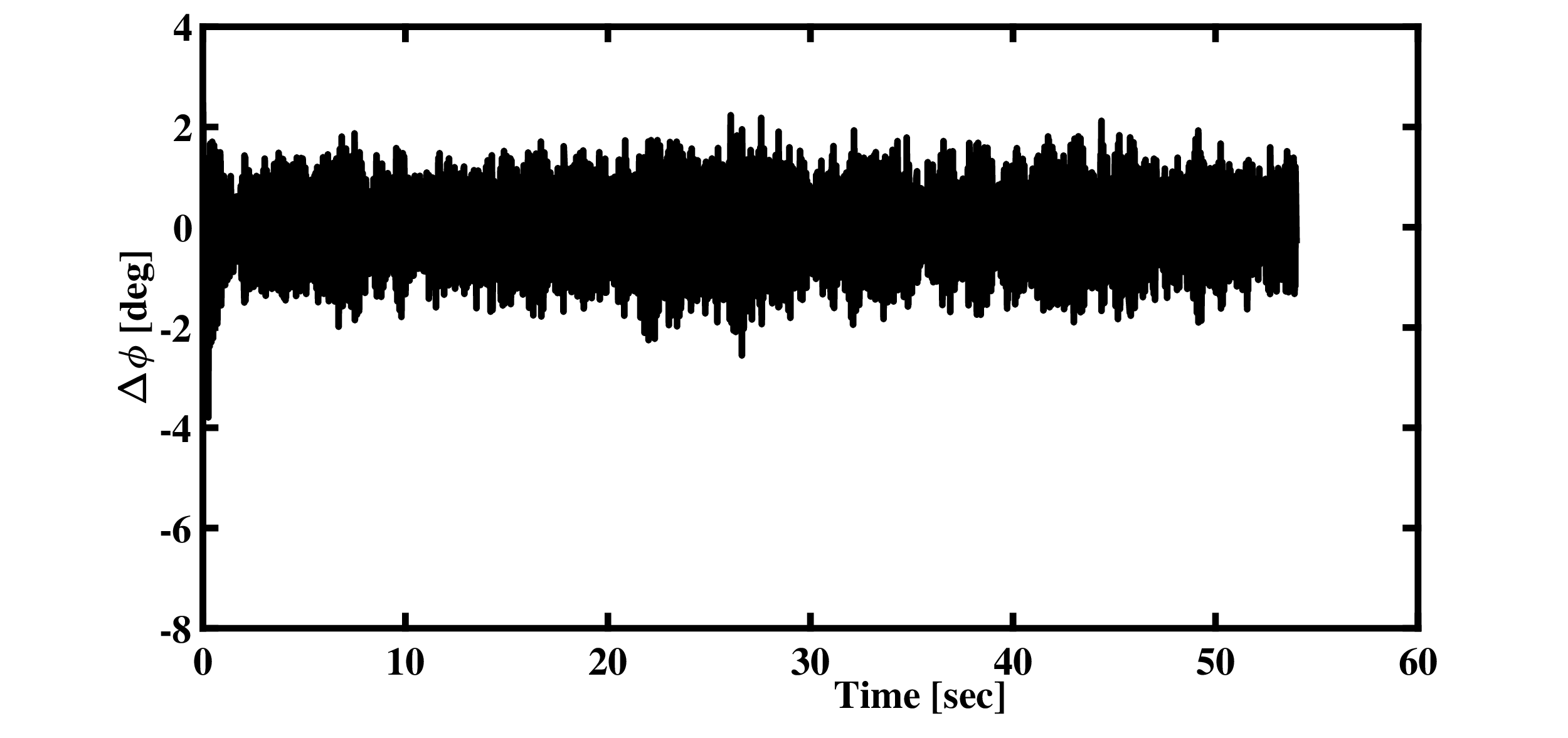}
        \caption{Carrier phase error (moderate)}
    \end{subfigure}
    \begin{subfigure}{0.3\linewidth}
        \includegraphics[width=\linewidth]{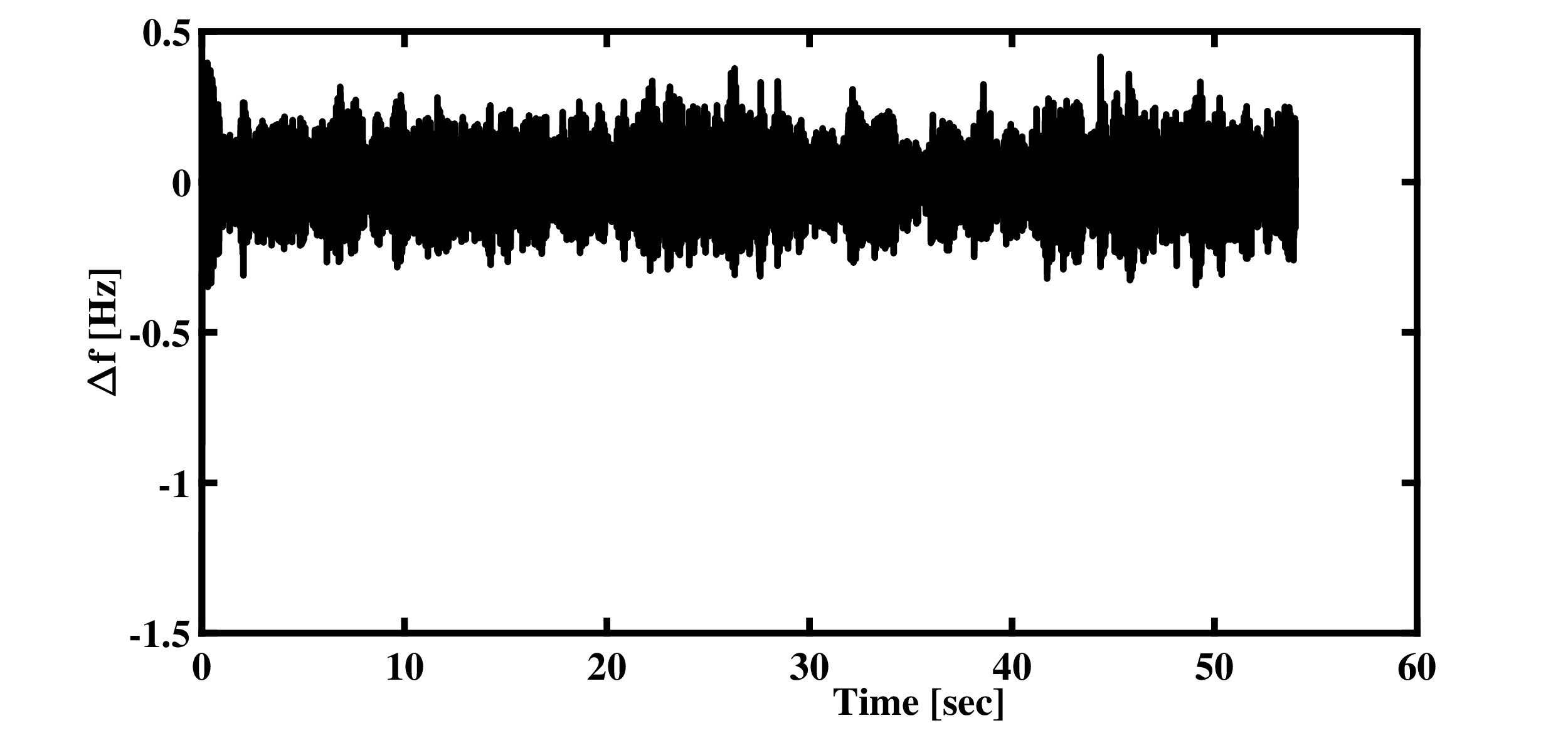}
        \caption{Carrier frequency error (moderate)}
    \end{subfigure}
     \vfill
    \begin{subfigure}{0.3\linewidth}
        \includegraphics[width=\linewidth]{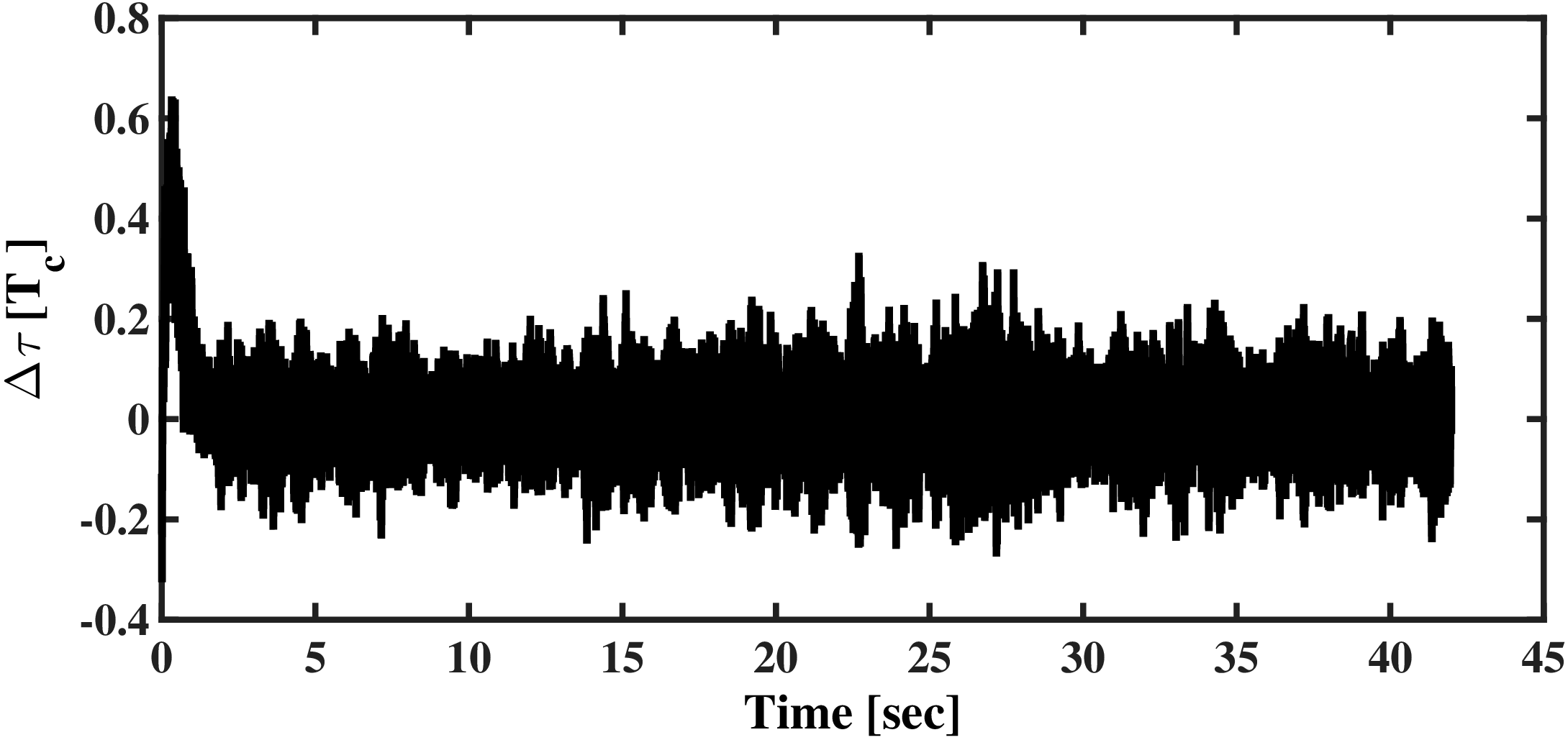}
        \caption{Code phase error (High dynamic)}
    \end{subfigure}
    \begin{subfigure}{0.3\linewidth}
        \includegraphics[width=\linewidth]{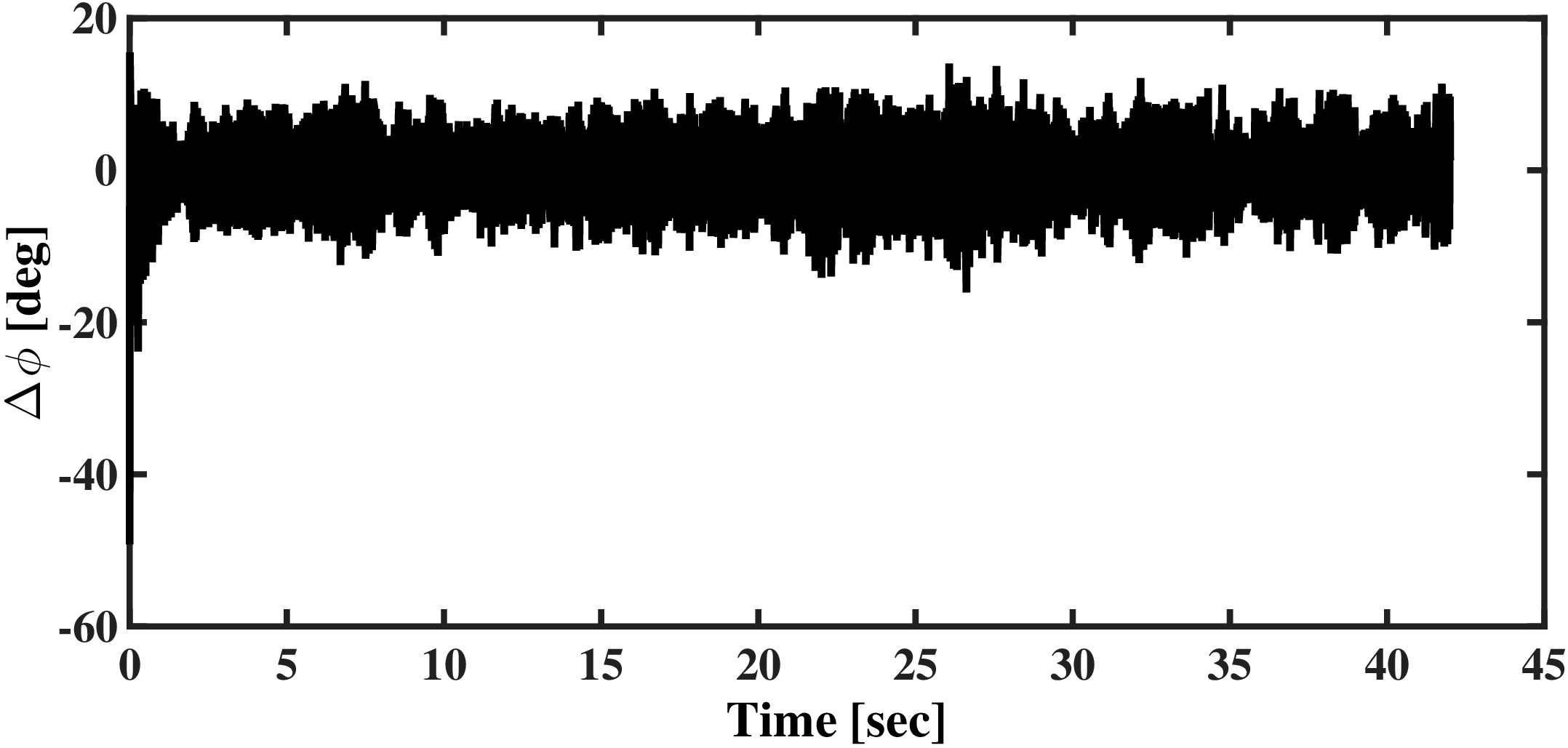}
        \caption{Carrier phase error (High dynamic)}
    \end{subfigure}
    \begin{subfigure}{0.3\linewidth}
        \includegraphics[width=\linewidth]{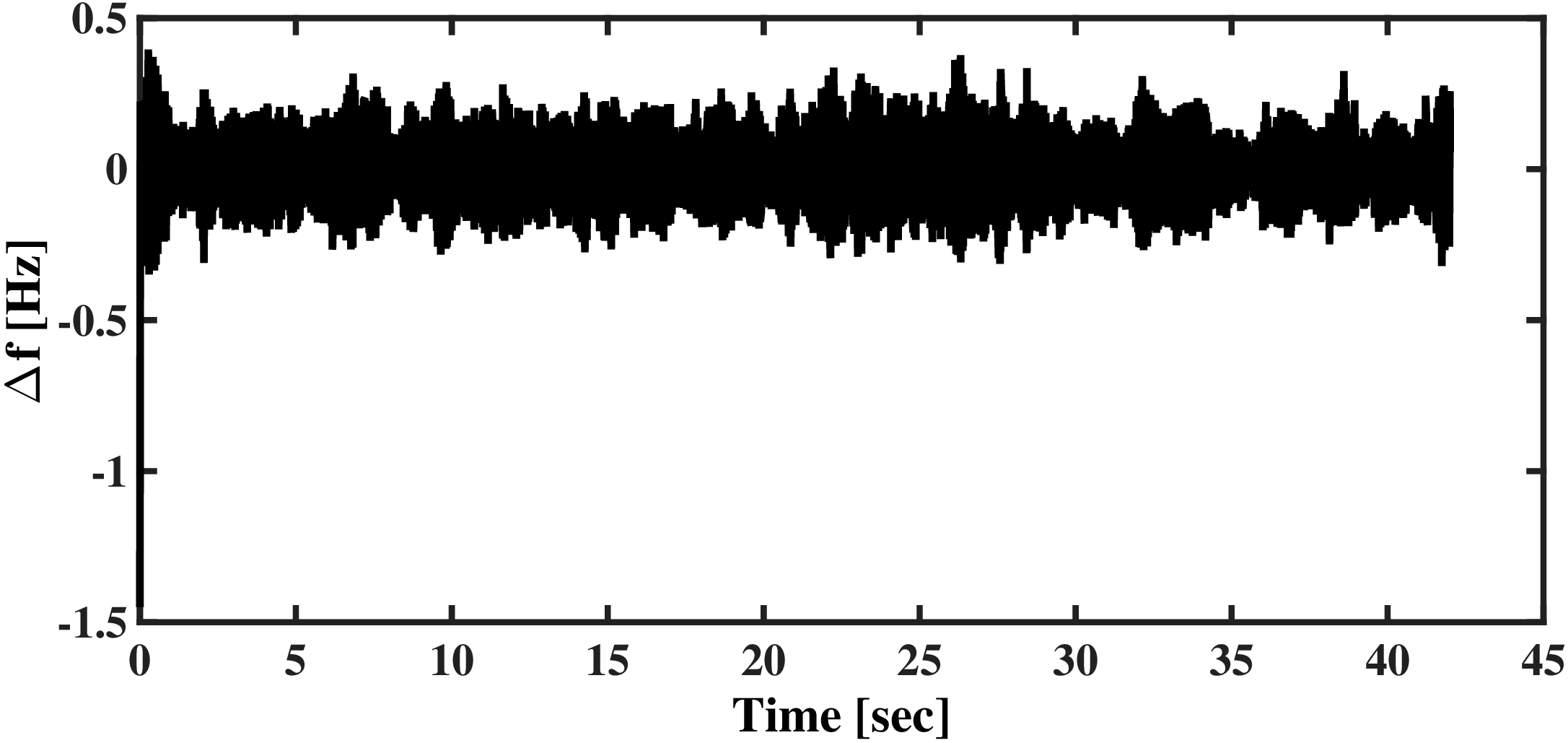}
        \caption{Carrier frequency error (High dynamic)}     
    \end{subfigure}
    \caption{Tracking‑loop discriminator outputs for PRN~26 under static, moderate and dynamic scenarios.}
    \label{fig:discriminator}
\end{figure*}

To assess tracking performance, Fig.~\ref{fig:discriminator} presents discriminator outputs for the DLL, PLL, and FLL loops under static and dynamic conditions. In all cases, the code‑phase, carrier‑phase, and carrier‑frequency errors remain well within the expected bounds for stable tracking \cite{kaplanHegarty}. These results indicate that the injected code‑phase and Doppler dynamics are consistent with the receiver’s internal models, allowing the tracking loops to maintain lock even under rapid motion.

\begin{figure*}[t]
    \centering
    \begin{subfigure}{0.3\linewidth}
        \includegraphics[width=\linewidth]{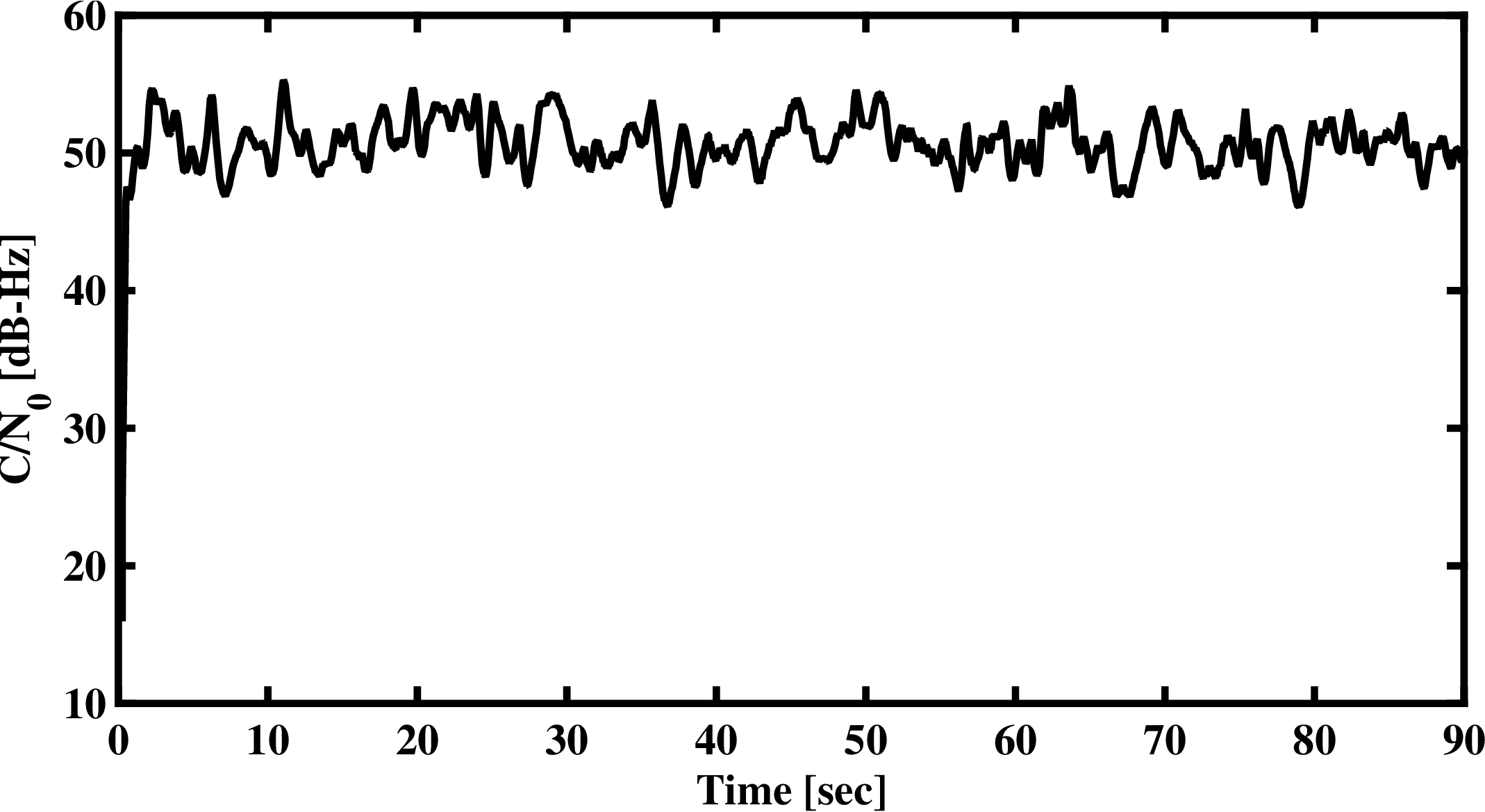}
        \caption{Static scenario}
    \end{subfigure}
    \begin{subfigure}{0.3\linewidth}
        \includegraphics[width=\linewidth]{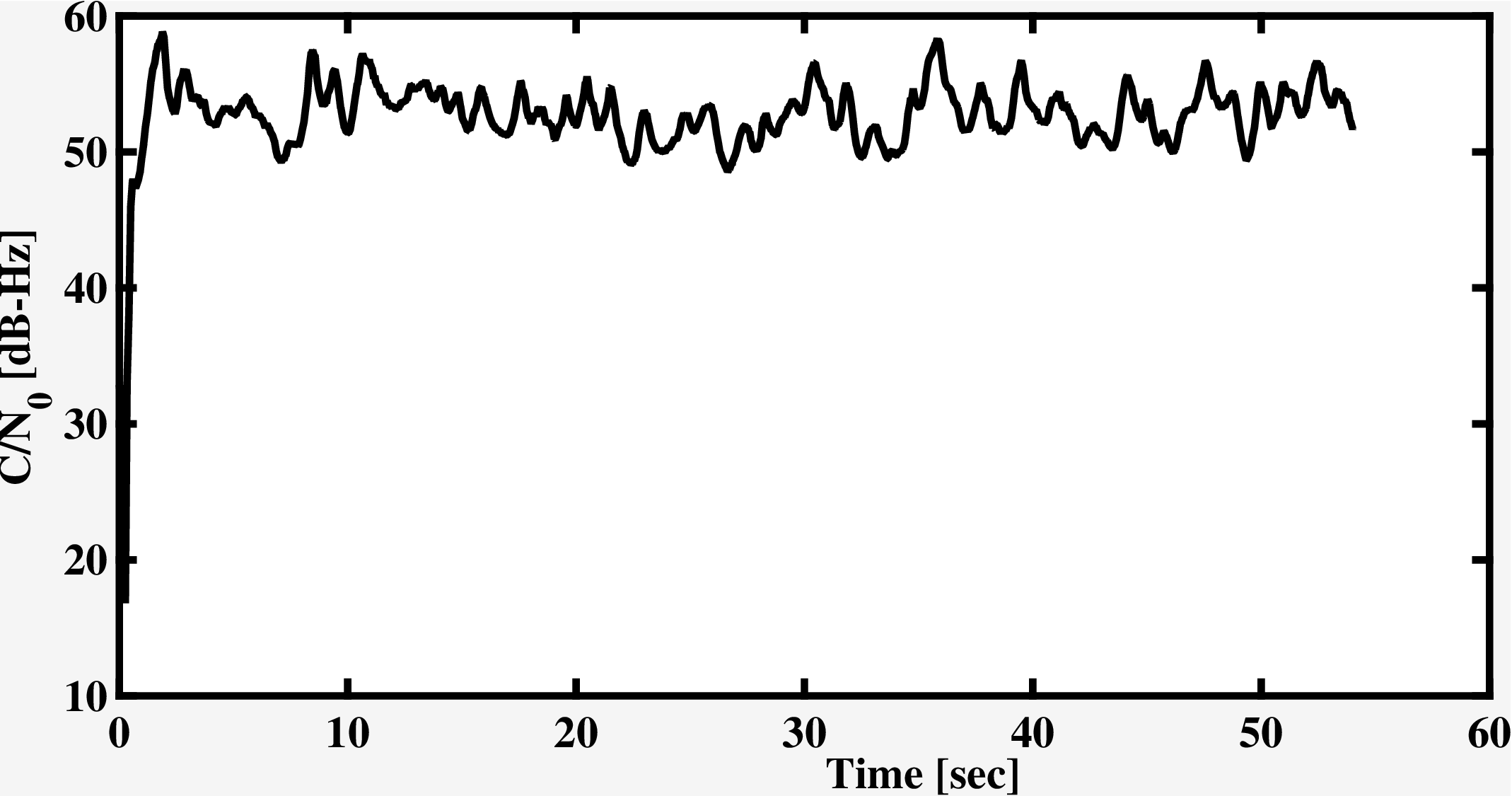}
        \caption{Moderate scenario}
    \end{subfigure}
    \begin{subfigure}{0.3\linewidth}
        \includegraphics[width=\linewidth]{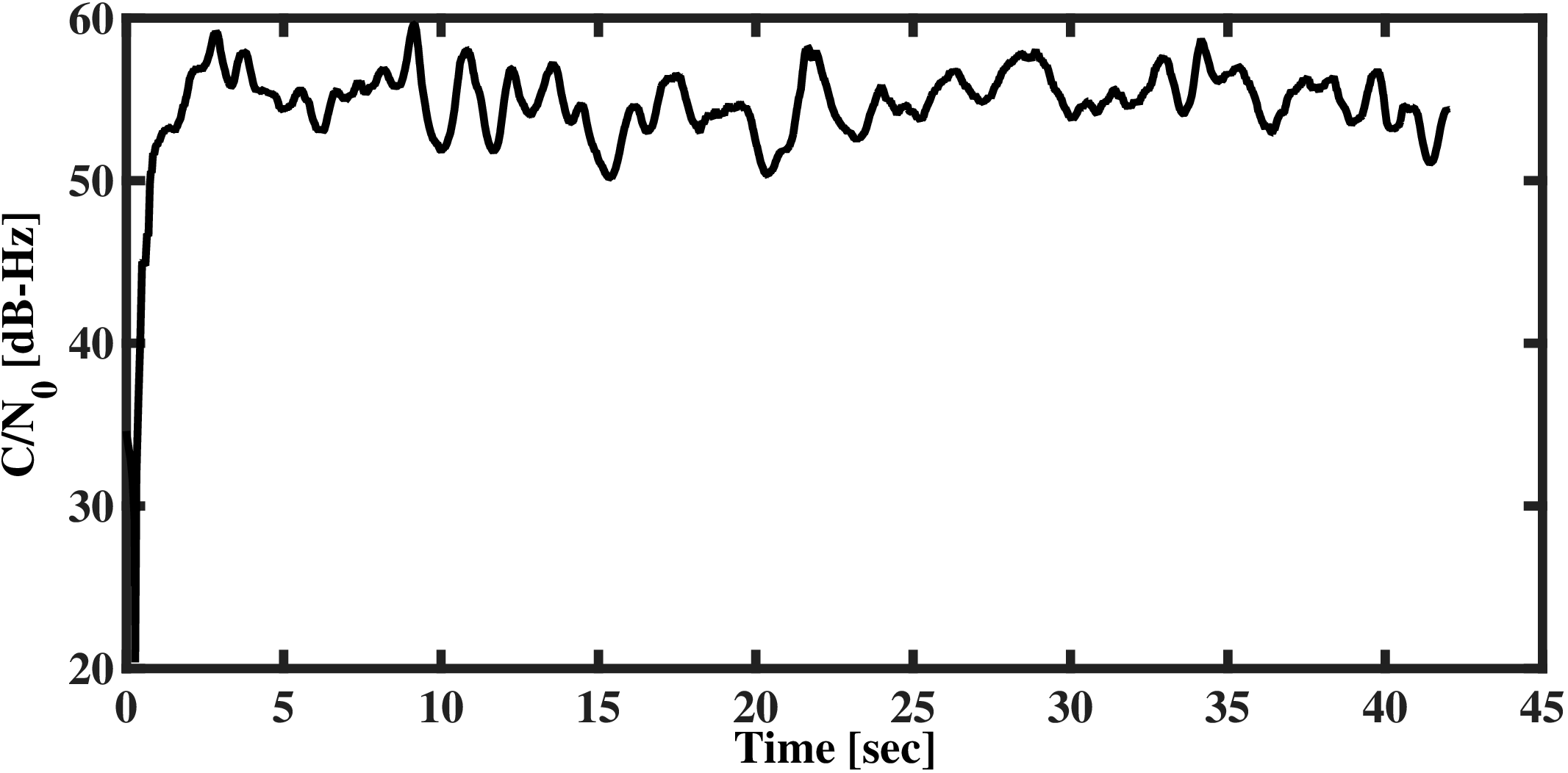}
        \caption{Higy dynamic scenario}
    \end{subfigure}    
    \caption{Carrier‑to‑noise density ratio $(C/N_0)$ for PRN~26.}
    \label{fig:CNR}
\end{figure*}

Figure~\ref{fig:CNR} shows the measured carrier‑to‑noise density ratio $(C/N_0)$ for PRN~26. The observed values, typically between 40–50~dB‑Hz, are consistent with open‑sky conditions and the simulated satellite geometry. Because the digital twin explicitly models received power, $C/N_0$ can be varied in a controlled manner to evaluate receiver sensitivity and robustness.

\begin{figure}[t]
    \centering
    \includegraphics[width=1.3\linewidth]{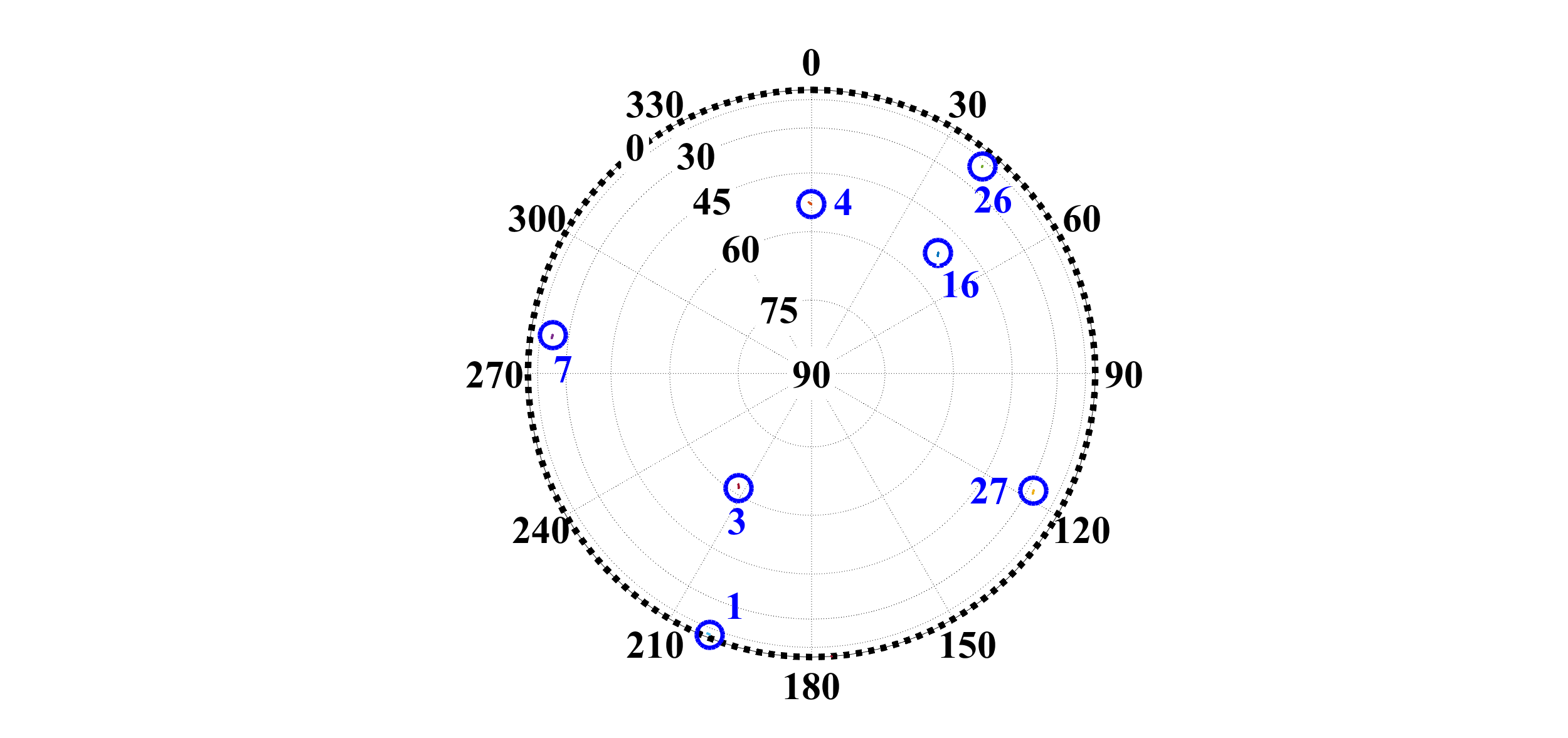}
    \caption{Sky plot of satellites used in the PVT solution.}
    \label{fig:skyplot}
\end{figure}

\begin{figure*}[t]
    \centering
    \begin{subfigure}{0.45\linewidth}
        \includegraphics[width=\linewidth]{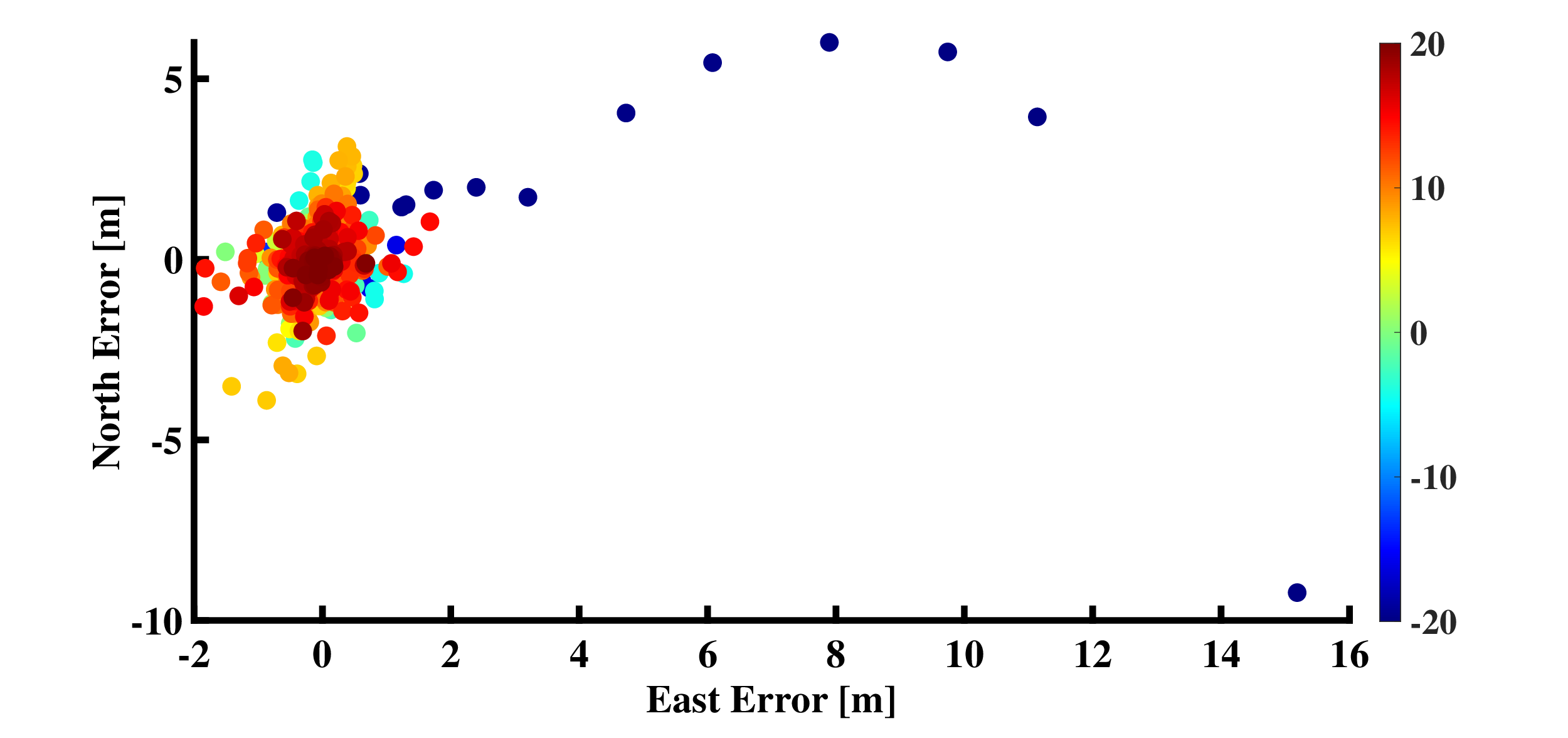}
        \caption{Horizontal position error (static)}
        \label{fig:HorErrstatic}
    \end{subfigure}
    \begin{subfigure}{0.45\linewidth}
        \includegraphics[width=\linewidth]{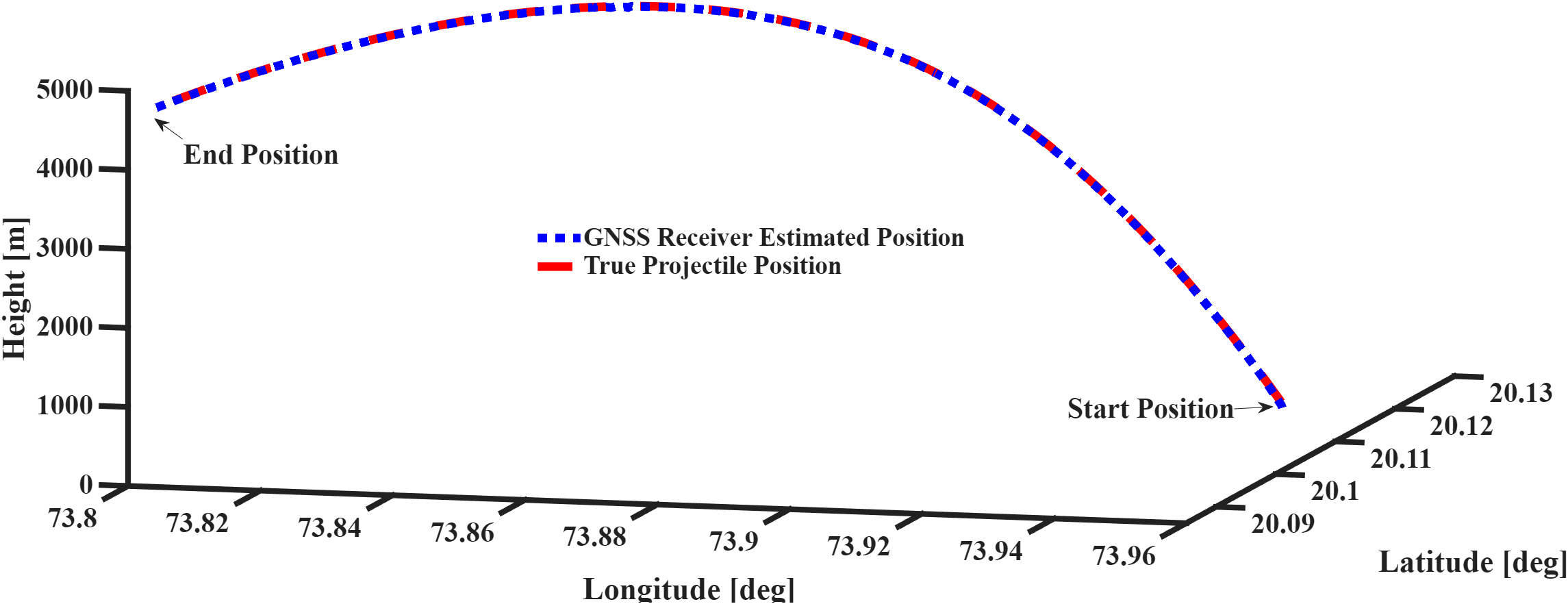}
        \caption{Estimated vs.\ true projectile trajectory (dynamic)}
        \label{fig:3DTrajectdynamic}
    \end{subfigure}
    \caption{Position‑domain validation for static and high‑dynamics scenarios.}
    \label{fig:PosPerf}
\end{figure*}

Position‑domain results further validate the physical consistency of the synthesized signal. Figure~\ref{fig:skyplot} shows the sky plot of satellites used in the PVT solution. For the static case, the horizontal position error (Fig.~\ref{fig:HorErrstatic}) remains within $\pm 2$~m, demonstrating that the digital twin produces stable and accurate pseudorange measurements. For the high‑dynamics scenario, Fig.~\ref{fig:3DTrajectdynamic} compares the software receiver’s estimated trajectory with the truth‑model projectile trajectory. The close alignment between the two confirms that the injected Doppler and Doppler‑rate dynamics remain consistent even under rapid motion.

\begin{figure*}[t]
    \centering
    \includegraphics[width=\textwidth]{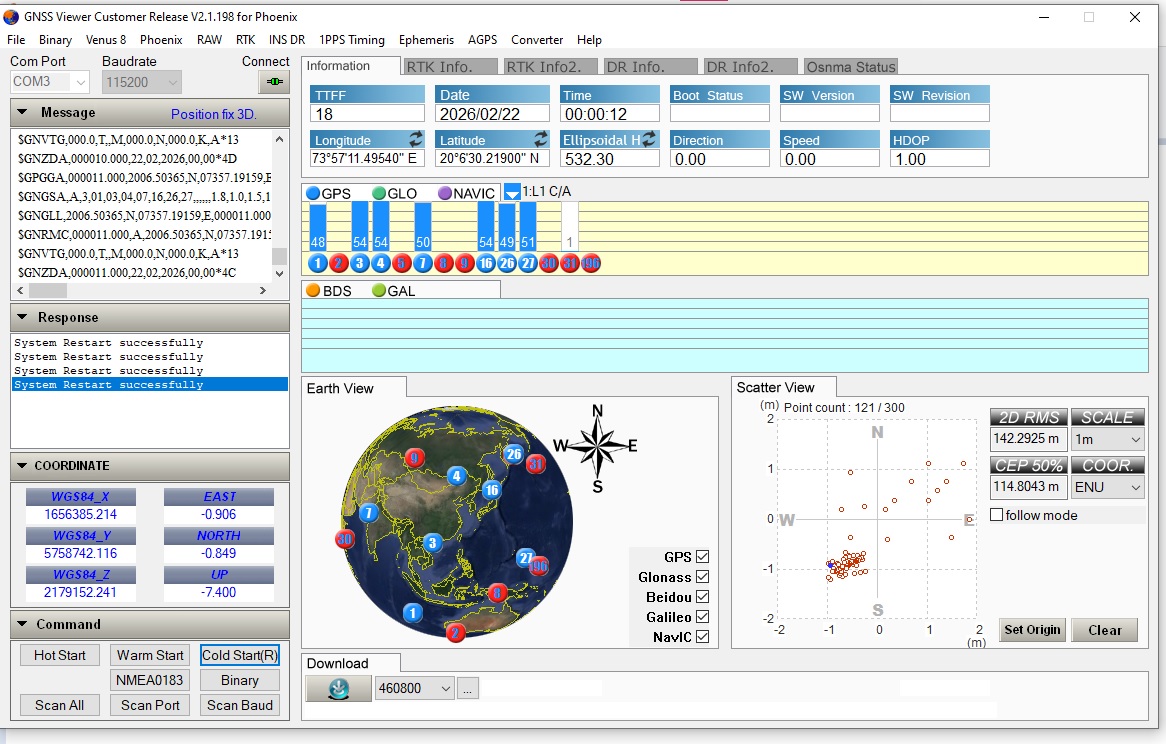}
    \caption{HIL results from the SkyTraq PX1125S‑01A receiver, showing satellite tracking, signal strength, and horizontal scatter plot.}
    \label{fig:skyTraq}
\end{figure*}

Finally, Fig.~\ref{fig:skyTraq} presents results from the SkyTraq PX1125S‑01A receiver operating in HIL mode. The scatter plot shows horizontal deviations within approximately $\pm 2$~m, closely matching the software receiver’s performance. This agreement demonstrates that the digital twin not only produces physically consistent IF samples but also generates RF signals that commercial receivers can track reliably.

\subsection{Receiver Clock Bias and Drift Validation}

In addition to validating code, carrier, and Doppler consistency, the stability of the receiver clock provides an important indicator of the temporal fidelity of the synthesized signal. Because the digital twin injects Doppler and Doppler-rate dynamics derived directly from satellite ephemerides and the user trajectory, the receiver’s estimated clock bias and drift should evolve smoothly and remain consistent with the truth-model timing behavior, without exhibiting spurious jumps or oscillations.

Clock stability is assessed using the time-bias estimates produced by the PVT solution. The analysis proceeds in two stages. First, the clock bias is logged as a time series and a linear trend is fitted to extract the average bias and drift rate. The slope of this trend represents the fractional frequency offset of the receiver oscillator: a small slope indicates a stable clock, whereas a larger slope reflects increased drift. Second, short-term stability is quantified using the Allan deviation, which is computed from the time-offset sequence following \cite{allan1987time}. The fractional frequency deviation is obtained as
\begin{equation}
    \tilde{\delta t}_k = \frac{\delta t_{k+1} - \delta t_k}{\tau_0},
\end{equation}
where $\tau_0$ is the sampling interval. The Allan deviation is then
\begin{equation}
    \sigma_{\tilde{\delta t}}(\tau) = \sqrt{\frac{1}{2(N-1)} \sum_{i=1}^{N-1} \left( \tilde{\delta t}_{i+1} - \tilde{\delta t}_{i} \right)^2 },
\end{equation}
with $N$ denoting the number of samples.

\begin{figure}[t]
    \centering
    \includegraphics[width=1\linewidth]{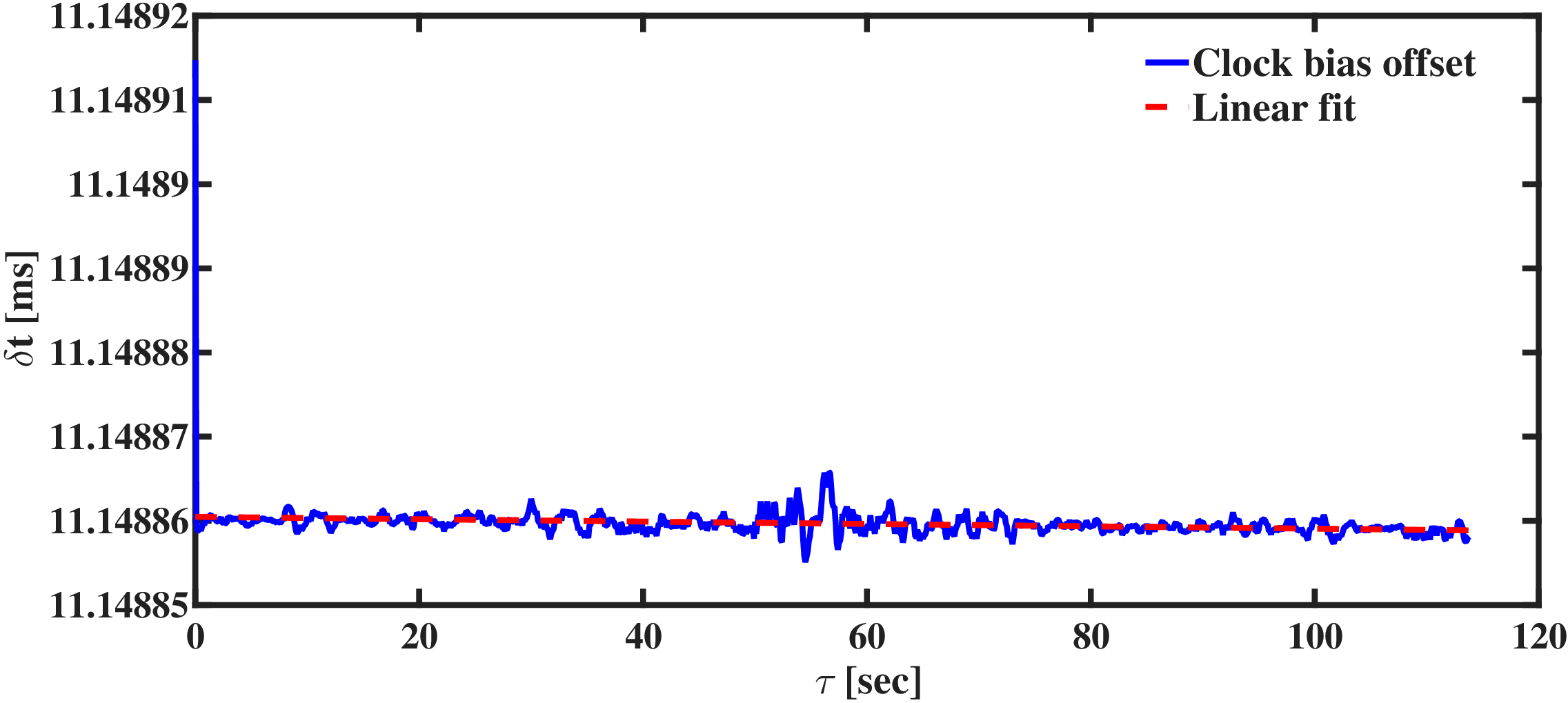}
    \caption{Receiver clock-bias time series and linear drift estimate (static scenario).}
    \label{fig:clockBiasTimeSeries}
\end{figure}

Figure~\ref{fig:clockBiasTimeSeries} shows the estimated clock-bias time series for a 120-second static experiment. The bias remains tightly bounded between 11.14885 and 11.14892~ms, with only small fluctuations attributable to oscillator noise and measurement uncertainty. The superimposed linear fit provides the drift estimate and confirms that the receiver clock evolves smoothly over time, without discontinuities or artifacts introduced by the digital-twin signal generation or the SDR transmission chain.

\begin{figure}[t]
    \centering
    \includegraphics[width=1\linewidth]{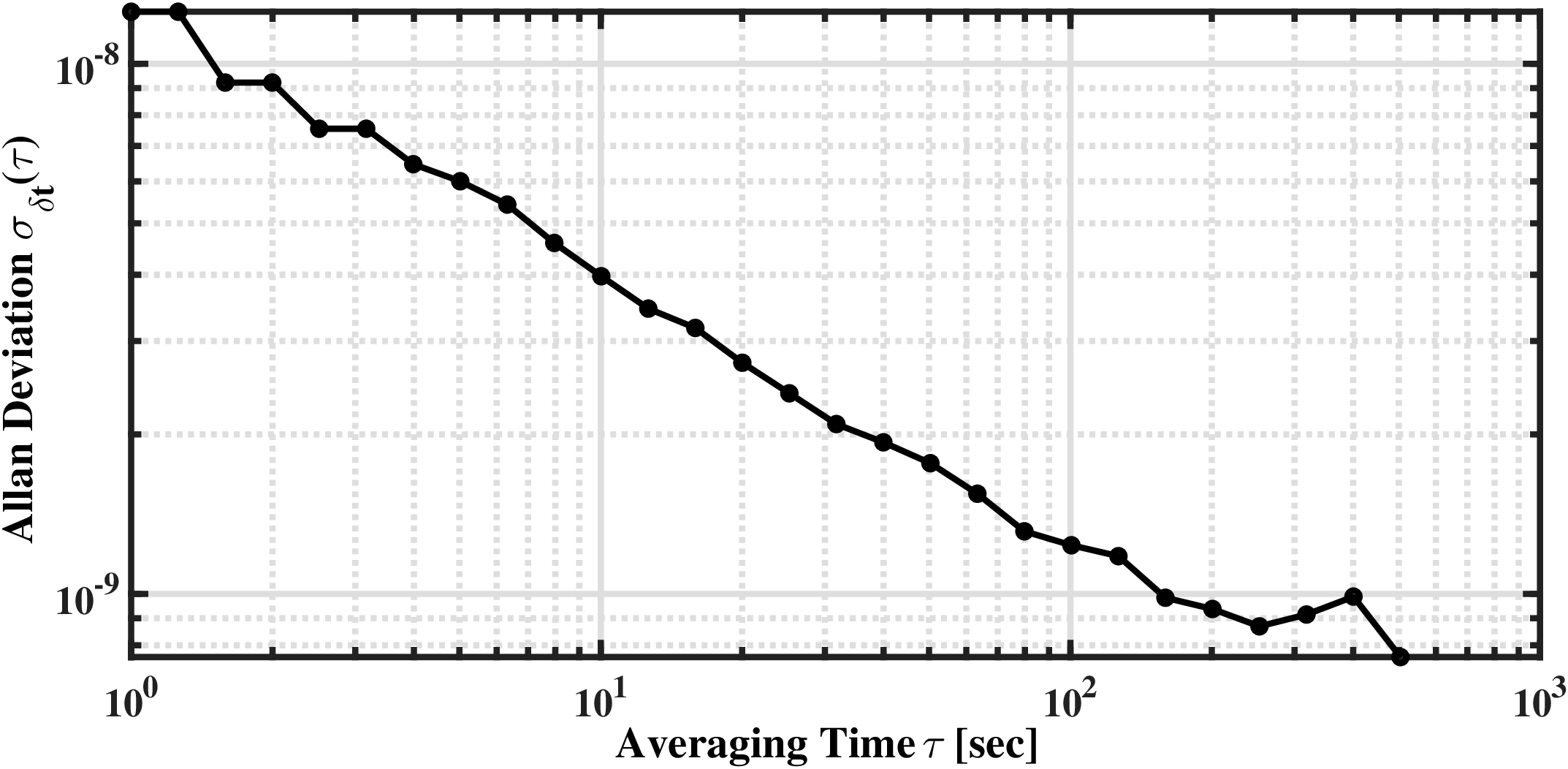}
    \caption{Allan deviation of the receiver clock derived from the clock-bias sequence.}
    \label{fig:AllanDev}
\end{figure}

Figure~\ref{fig:AllanDev} presents the corresponding Allan deviation. For short averaging intervals ($\tau \approx 1$--$100$~s), the deviation is relatively high, reflecting short-term jitter and white-noise behavior typical of quartz oscillators in GNSS receivers. As $\tau$ increases, the Allan deviation decreases and stabilizes in the $10^{-9}$--$10^{-8}$ range, consistent with the expected performance of low-cost receiver oscillators. The absence of anomalous peaks or irregularities indicates that the digital-twin-generated Doppler and timing dynamics do not introduce artificial frequency perturbations.

Overall, the clock-domain results confirm that the synthesized signal preserves the temporal coherence required for reliable PVT estimation. The SDR-based RF generation does not introduce additional timing artifacts, and the receiver’s estimated clock behavior remains physically consistent with the truth-model dynamics. This validation complements the code, carrier, and position-domain analyses, providing an additional layer of evidence for the fidelity of the proposed digital twin.

\subsection{Tracking Loop Jitter Analysis Under High Dynamics}

Tracking loop stability under rapid motion is further evaluated by analyzing the jitter in the DLL, PLL, and FLL discriminator outputs. Tracking jitter reflects the receiver’s ability to follow fast variations in code phase, carrier phase, and Doppler frequency, and therefore provides a sensitive indicator of signal fidelity—particularly in high-dynamics conditions. Because closed-loop tracking behavior is nonlinear and difficult to characterize analytically, practical stability thresholds or ``rules of thumb'' are commonly used to assess robustness.

\textbf{Code Tracking (DLL):}
In the absence of multipath, DLL jitter is dominated by thermal noise and dynamic stress. A widely used guideline requires the $3\sigma_{\mathrm{DLL}}$ code-phase error to remain within half the linear pull-in range of the discriminator~\cite{kaplanHegarty}:
\[
3\sigma_{\mathrm{DLL}} \le \frac{d}{2},
\]
where $d$ is the early–late correlator spacing. For $d = 0.5$ chips, the corresponding threshold is $\sigma_{\mathrm{DLL,Th}} = \tfrac{1}{12}$ chips.

\textbf{Carrier Tracking (PLL):}
PLL stability is governed by thermal noise, oscillator imperfections, and dynamic stress. For an arctangent Costas discriminator, the $3\sigma_{\mathrm{PLL}}$ phase error must remain within one-quarter of the $180^\circ$ pull-in range~\cite{kaplanHegarty}, yielding a threshold of $\sigma_{\mathrm{PLL,Th}} = 15^\circ$.

\textbf{Frequency Tracking (FLL):}
For the FLL, the rule of thumb requires the $3\sigma_{\mathrm{FLL}}$ frequency error to remain within one-quarter of the discriminator pull-in range $(\kappa)$~\cite{kaplanHegarty}. With $\kappa = \pm \tfrac{1}{2T}$ for integration time $T$, the corresponding threshold is $\sigma_{\mathrm{FLL,Th}} = \tfrac{1}{12T}$.

These thresholds represent the maximum allowable standard deviation of the tracking error; exceeding them typically results in loss of lock or cycle slips. Robust tracking requires the measured jitter $\sigma$ to remain below the theoretical limit $\sigma_{\mathrm{Th}}$.

\begin{figure*}[t]
    \centering
    \begin{subfigure}{0.32\linewidth}
        \includegraphics[width=\linewidth]{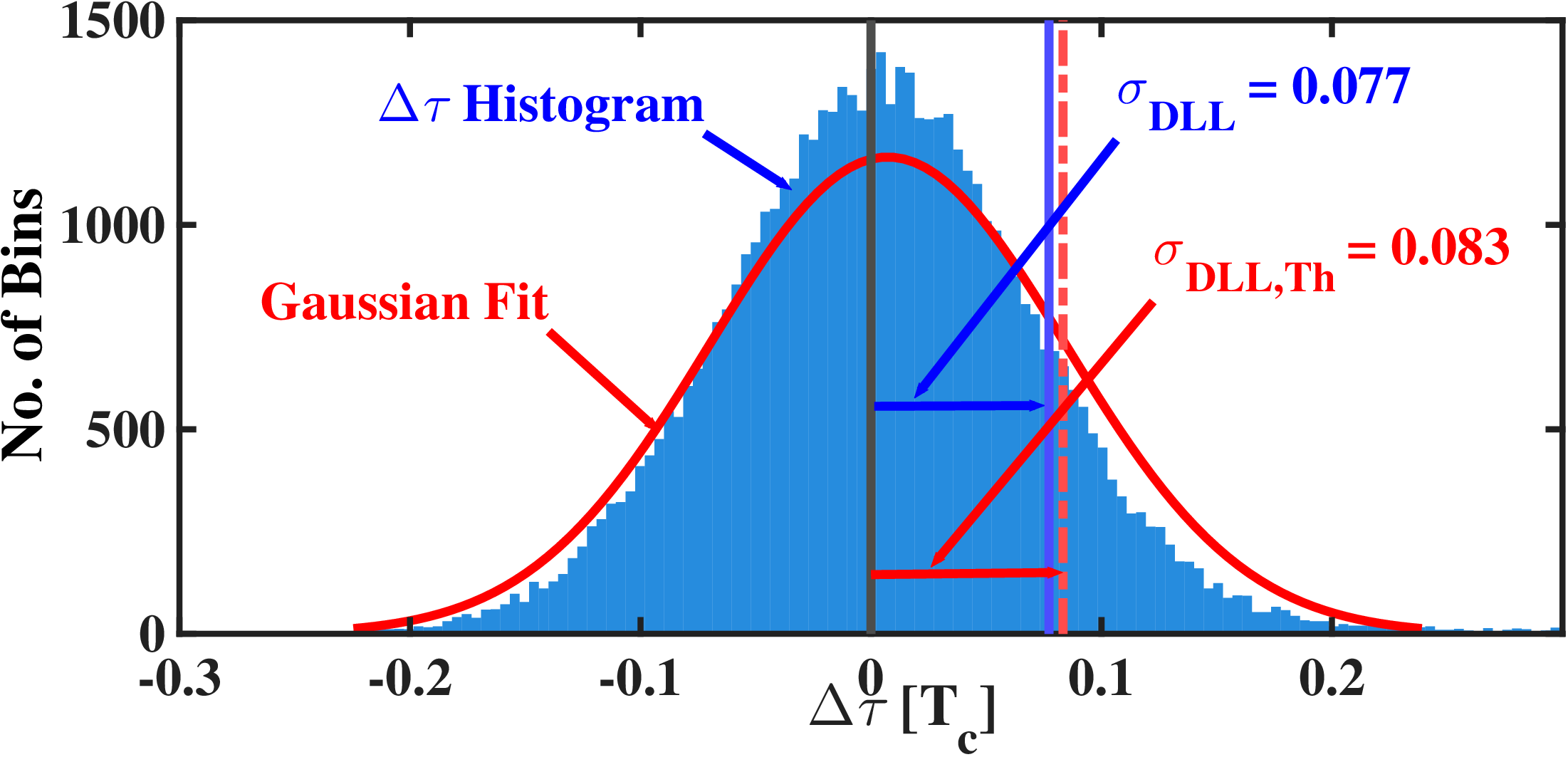}
        \caption{DLL jitter distribution}
        \label{fig:DLL_Err}
    \end{subfigure}
    \begin{subfigure}{0.32\linewidth}
        \includegraphics[width=\linewidth]{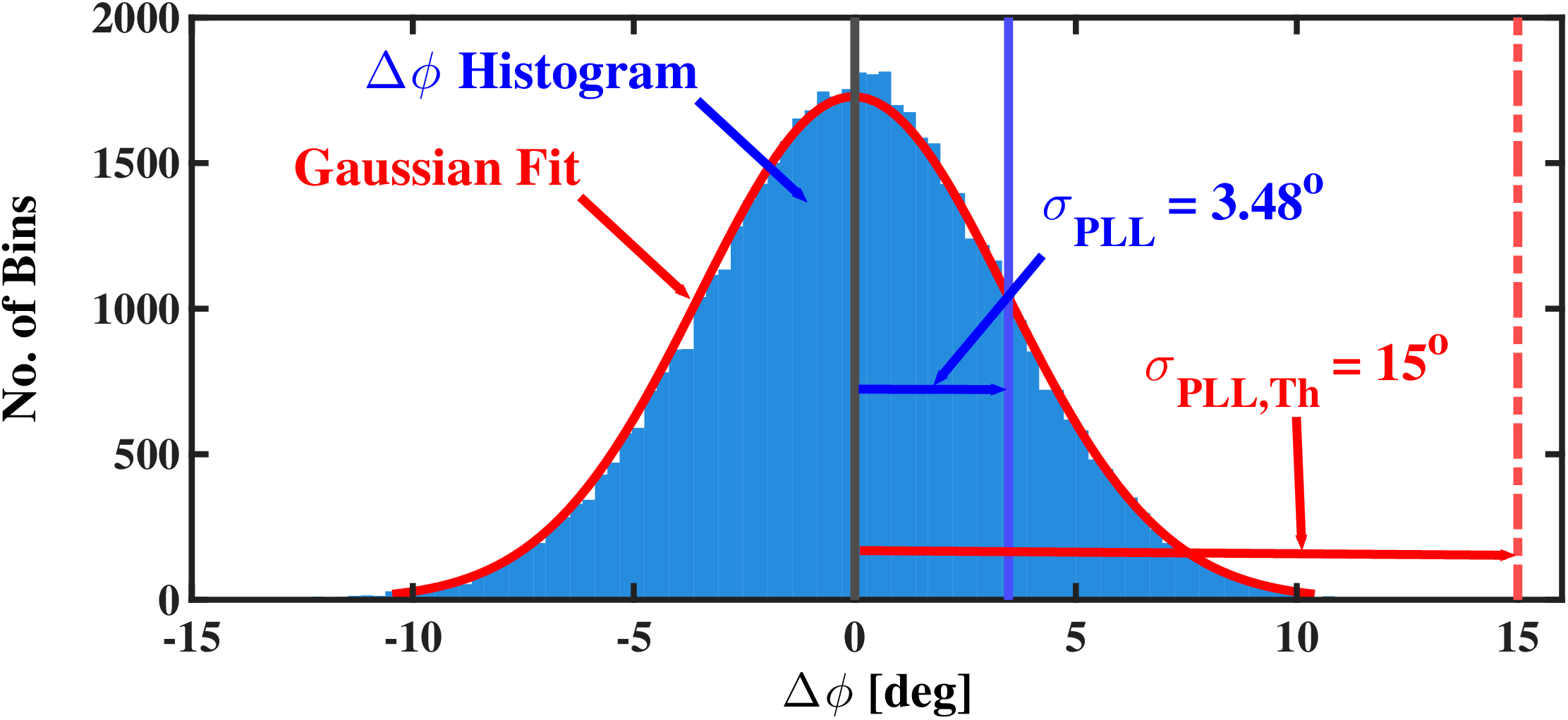}
        \caption{PLL jitter distribution}
        \label{fig:PLL_Err}
    \end{subfigure}
    \begin{subfigure}{0.32\linewidth}
        \includegraphics[width=\linewidth]{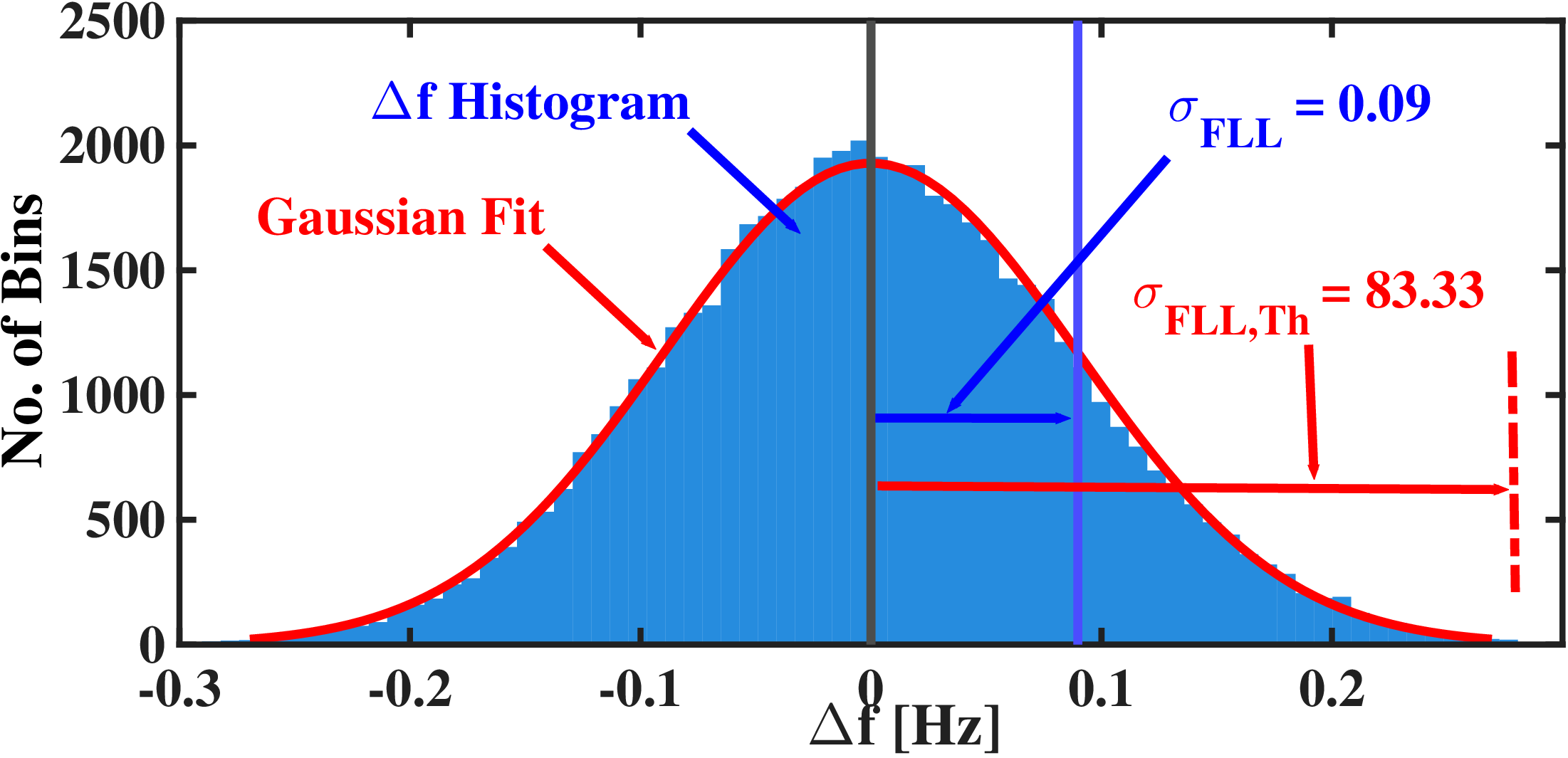}
        \caption{FLL jitter distribution}
        \label{fig:FLL_Err}
    \end{subfigure}
    \caption{Measured and theoretical standard deviations of DLL, PLL, and FLL tracking errors for PRN~26 under high-dynamics motion.}
    \label{fig:jitterDynamic}
\end{figure*}

Figure~\ref{fig:jitterDynamic} presents the measured jitter distributions for PRN~26 under high-dynamics motion. The DLL histogram in Fig.~\ref{fig:DLL_Err} shows a measured standard deviation of $\sigma_{\mathrm{DLL}} = 0.077$ chips, which lies just below the theoretical limit of $\sigma_{\mathrm{DLL,Th}} = 0.083$ chips. Although the loop remains stable, the narrow margin indicates that the DLL is operating near its stress boundary, consistent with the rapid code-phase variations induced by the high-acceleration trajectory.

The PLL results in Fig.~\ref{fig:PLL_Err} show a measured jitter of $\sigma_{\mathrm{PLL}} = 3.48^\circ$, significantly below the $15^\circ$ threshold. This large safety margin indicates that the carrier-phase tracking loop remains highly stable, with no risk of cycle slips even during the most demanding portions of the trajectory.

The FLL histogram in Fig.~\ref{fig:FLL_Err} shows a measured jitter of $\sigma_{\mathrm{FLL}} = 0.09$~Hz, compared to a theoretical limit of $\sigma_{\mathrm{FLL,Th}} = 83.33$~Hz. This represents an extremely wide stability margin, confirming that the injected Doppler and Doppler-rate dynamics are smooth and physically consistent, and that the FLL is virtually unaffected by the high-dynamics conditions.

Overall, all three tracking loops maintain lock throughout the high-dynamics scenario, with measured jitter remaining below the corresponding theoretical thresholds. However, the results also highlight that the DLL is the most vulnerable loop under rapid motion, as its jitter approaches the pull-in boundary. In contrast, the PLL and FLL exhibit substantial robustness margins. These findings demonstrate that the digital twin accurately reproduces both the mean behavior and higher-order temporal variations of the GNSS observables, enabling realistic stress testing of receiver tracking loops under extreme dynamics.

\section{CONCLUSION}

This paper presented a physics‑driven digital twin framework for the GPS L1 C/A signal chain, enabling end‑to‑end modeling, hardware‑in‑the‑loop testing, and reproducible evaluation of GNSS receiver performance. The framework generates complex baseband signals by injecting trajectory‑consistent code‑phase, Doppler, and Doppler‑rate dynamics derived directly from satellite ephemerides and user motion. By incorporating ionospheric and tropospheric delay models, the digital twin recreates realistic propagation conditions and provides a controlled environment for exercising receiver acquisition, tracking, and navigation algorithms.

Experimental validation was carried out across static, moderate‑motion, and high‑dynamics scenarios using both simulated IF samples and SDR‑recorded data. The close agreement between truth‑model observables and receiver‑estimated pseudorange, Doppler, and carrier‑phase measurements confirms the fidelity of the injected dynamics. The strong correspondence between simulated and measured IF signals further verifies the accuracy of the SDR‑based HIL implementation. Position‑domain results from both software and commercial receivers demonstrate meter‑level consistency, reinforcing the physical realism of the synthesized signal.

Overall, the proposed digital twin provides a high‑fidelity, repeatable, and flexible platform for GNSS receiver development, algorithm evaluation, and performance benchmarking. Future extensions will target multi‑frequency and multi‑constellation operation (e.g., NavIC, BDS, GLONASS), integration of advanced interference and spoofing models, and real‑time closed‑loop testing with embedded navigation and guidance systems.

\section*{ACKNOWLEDGMENT}

The authors used Microsoft Copilot to improve the grammar and clarity of the manuscript. AI assistance was limited strictly to language refinement; all technical content, analyses, and conclusions were fully authored, validated, and verified by the authors.

\end{document}